\documentclass[10pt,conference]{IEEEtran}
\IEEEoverridecommandlockouts

\usepackage{multirow}
\usepackage{subcaption}
\usepackage{enumitem}
\usepackage{algorithm}
\usepackage[ruled,linesnumbered,algo2e]{algorithm2e}
\usepackage{tcolorbox}
\usepackage{listings}
\usepackage{booktabs}
\usepackage{graphicx} 
\usepackage{bm}
\usepackage{xspace}
\usepackage{caption}
\usepackage{url}
\usepackage{amsmath,amssymb,amsfonts}
\usepackage{textcomp}
\usepackage{xcolor}
\usepackage{cite}
\usepackage{tcolorbox}
\usepackage{color,xcolor,colortbl}
\usepackage{listings}  
\def\BibTeX{{\rm B\kern-.05em{\sc i\kern-.025em b}\kern-.08em
    T\kern-.1667em\lower.7ex\hbox{E}\kern-.125emX}}

\DontPrintSemicolon
\SetAlgoNlRelativeSize{0}
\SetAlgoNlRelativeSize{-1}
\newcommand{\resetalgocf}{\setcounter{AlgoLine}{0}}

\definecolor{mygray}{gray}{.9}
\newcommand{\g}{\cellcolor{mygray}}
\newcommand{\tool}{SBLLM\xspace}
\newcommand{\tabincell}[2]{\begin{tabular}{@{}#1@{}}#2\end{tabular}}

\usepackage{tikz}
\newcommand*{\circled}[1]{\lower.7ex\hbox{\tikz\draw (0pt, 0pt)%
    circle (.5em) node {\makebox[1em][c]{\small #1}};}}

\begin{document}

\title{Search-Based LLMs for Code Optimization}

\author{\IEEEauthorblockN{Shuzheng Gao$^{1}$, Cuiyun Gao$^{2\ast}$, Wenchao Gu$^{1}$, Michael R. Lyu$^{1}$}

\IEEEauthorblockA{$^1$ Department of Computer Science and Engineering, The Chinese University of Hong Kong, China}

\IEEEauthorblockA{$^2$ School of Computer Science and Technology, Harbin Institute of Technology, Shenzhen, China}

\IEEEauthorblockA{szgao23@cse.cuhk.edu.hk, gaocuiyun@hit.edu.cn, wcgu@cse.cuhk.edu.hk, lyu@cse.cuhk.edu.hk}

\thanks{$^{\ast}$ Corresponding author. The author is also affiliated with Peng Cheng Laboratory.}}

\maketitle

\begin{abstract}
The code written by developers usually
suffers from efficiency problems and contain various performance bugs. These inefficiencies necessitate the research of automated refactoring methods for code optimization. Early research in code optimization
employs
rule-based methods and focuses
on specific inefficiency issues,
which are labor-intensive and suffer from the low coverage issue.
Recent work regards the task as a sequence generation problem, and resorts to deep learning (DL) techniques such as large language models (LLMs).
These methods typically prompt LLMs 
to directly generate optimized code.
Although these methods show state-of-the-art performance, such one-step generation paradigm is hard to achieve an optimal solution. 
First, complex optimization methods such as combinatorial ones are hard to be captured by LLMs. Second, the one-step generation paradigm poses challenge in precisely infusing the 
knowledge required for effective code optimization within LLMs,
resulting in 
under-optimized code.

To address these problems, we propose to model this task
from the search perspective, and propose a search-based LLMs framework named \tool that enables iterative refinement and discovery of improved optimization methods. \tool synergistically integrate LLMs with evolutionary search and 
consists of three key components: 1) an execution-based representative sample selection part that evaluates the fitness of each existing optimized code and prioritizes promising ones to 
pilot the generation of improved code;
2) an adaptive optimization pattern retrieval part that infuses targeted optimization patterns into the model for guiding LLMs towards rectifying and progressively enhancing their optimization methods;
and 3) a genetic operator-inspired chain-of-thought prompting part that aids LLMs in combining different optimization methods 
and generating
improved
optimization methods.
Our evaluation of \tool on a dataset of Python and C++ code demonstrates its effectiveness in improving code efficiency. Specifically, the results indicate that \tool can improve program execution efficiency by up to 209.59\% and consistently outperform all baseline methods by 8.75\% $\sim$ 28.06\% and 1.15\% $\sim$ 9.56\% with different LLMs in terms of top-5 speedup rate on Python and C++, respectively.

\end{abstract}

\pagestyle{plain}

\maketitle

\section{Introduction}\label{sec:intro} 


As referred in the ISO/IEC 25010 software quality guidelines, the computational efficiency of the software is a critical cornerstone of system performance and user satisfaction~\cite{ISOIEC,DBLP:conf/sigsoft/GargMCSW22}. 
Inefficient code snippets can induce increased system latency, computational resources waste, and lead to poor user experience, which is referred to as performance bugs~\cite{DBLP:conf/msr/NistorJT13,DBLP:conf/oopsla/JovicAH11}. 
Existing studies have demonstrated that these inefficiencies are widely {existed}
in software and are hard to detect and repair~\cite{DBLP:conf/oopsla/JovicAH11,DBLP:conf/cloud/DeanNGZRAJ14ta}. Consequently, {the task of code optimization, which aims to automatically refactor the code, simplify 
its
complexity and enhance performance metrics, has attracted researchers' attention in recent years.}

Early research in code optimization primarily focuses on rule-based methods, which {mainly} target specific types of inefficiencies {such as}
software misconfigurations~\cite{krishna2020cadet} and loop inefficiencies~\cite{DBLP:conf/msr/NistorJT13}. {These methods heavily rely on pre-defined rules created by experts, which are labor-intensive and suffer from the low coverage problem~\cite{DBLP:conf/sigsoft/GargMCSW22,DBLP:journals/corr/abs-2309-14846}.}
Recent advancements in deep learning (DL) {such as large language models (LLMs)} have inspired a burgeoning body of research.
These techniques learn code optimization patterns from data, broadening the array of inefficiency types that can be addressed.
{For example,} RapGen~\cite{DBLP:journals/corr/abs-2306-17077} introduces a retrieval-augmented generation approach with LLMs to
generate the optimized code 
in a zero-shot manner, surpassing the performance of previous fine-tuned small-sized 
{neural} models\cite{DBLP:conf/sigsoft/GargMCSW22}. Another recent work, PIE~\cite{DBLP:journals/corr/abs-2302-07867} establishes a benchmark with test cases 
and evaluates the performance of various 
prompting methods such as in-context learning (ICL)~\cite{DBLP:conf/nips/BrownMRSKDNSSAA20} and chain of thought (COT)~\cite{DBLP:journals/corr/abs-2201-11903}.

\begin{figure*}
    \centering
    \includegraphics[width=1\textwidth]{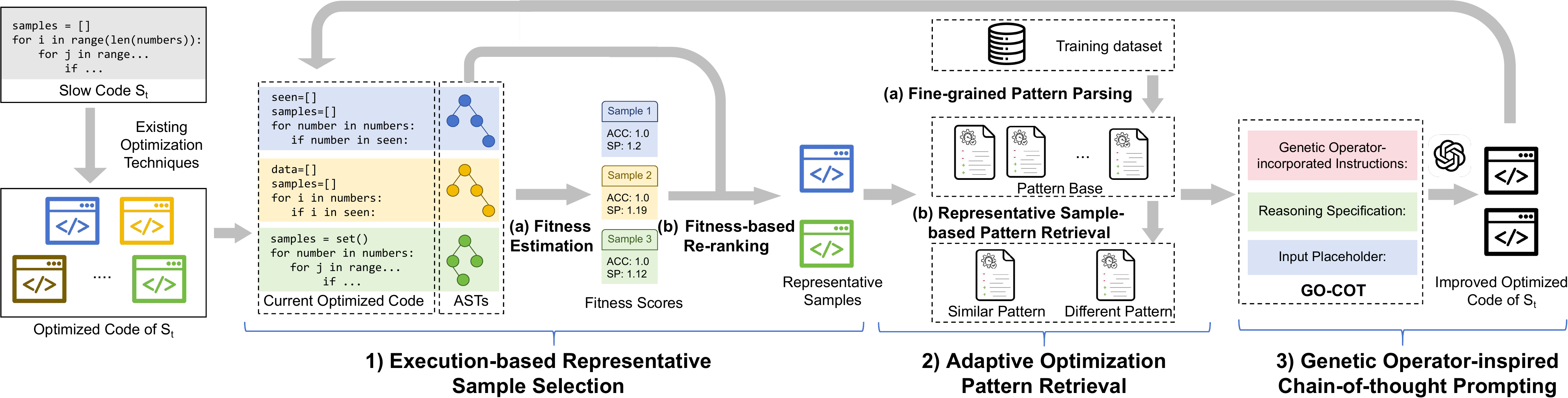}
    \caption{The overview of \tool.}
    \vspace{-0.4cm}
    \label{fig:overview}
\end{figure*}

Despite the success of LLM-based approaches, the adopted
one-step generation paradigm tends to largely limit the performance of code optimization for two main reasons.
First, it is challenging for LLMs to capture the complex optimization methods in one attempt. Code optimization involves a range of optimization levels~\cite{DBLP:journals/corr/abs-2402-07844}, from incremental improvements like removing some unnecessary computation to more substantial
optimizations that reshape
the whole algorithm. Moreover, optimization patterns are combinatorial in nature~\cite{DBLP:journals/smr/OuniKCSDI17,DBLP:journals/smr/GhannemEK14}, implying
that a single snippet may consist of multiple segments with the potential for different optimization.
Second, it is difficult to precisely integrate the essential 
knowledge required for effective code optimization into LLMs. Although some studies~\cite{DBLP:journals/corr/abs-2306-17077,DBLP:journals/jmlr/IzacardLLHPSDJRG23} have attempted to improve LLMs by retrieving similar code snippets, these approaches only consider the input similarity and ignores the characteristic of code optimization such as its combinatorial nature,
which may result in the generation of 
under-optimized code.


To mitigate the above challenges,
we propose to model 
the code optimization task from the search
perspective instead of the typical generation viewpoint. 
Specifically, the task of refactoring a given code snippet to be more efficient
can be formulated as a search problem, where the objective is to find the most efficient optimization method among a vast set of potential methods created by extensive code transformation methods and their possible combinations. 
In search-based software engineering, significant efforts have been devoted to advancing
the discovery of
optimal solutions in the large search space for various tasks~\cite{DBLP:journals/csur/HarmanMZ12,DBLP:journals/infsof/HarmanJ01}. 
These methods 
iteratively identify limitations in existing solutions to update
their strategy to propose better ones, aiming to progressively move towards the optimal solution~\cite{DBLP:conf/cav/0001LMN14,DBLP:journals/tse/GouesNFW12}.
Considering that a single generation by LLMs can be regarded as a one-step exploration in the search space,
the search-based approaches can benefit LLMs by incorporating them into the iterative refinement process.

In this work, we introduce a \textbf{S}earch-\textbf{B}ased \textbf{LLM}s for code optimization, named \tool. 
\tool synergistically combines LLMs and evolutionary search, comprising
three main components: 
1) Execution-based representative sample selection, where we leverage execution feedback to evaluate the fitness of each sample and prioritize representative ones with effective and distinct optimization methods to pilot further optimization. 
2) Adaptive optimization pattern retrieval, where we propose an adaptive retrieval mechanism to infuse domain knowledge in LLMs, and guide LLMs
to rectify and progressively enhance their optimization methods. 
3) Genetic operator-inspired chain-of-thought (GO-COT) prompting, where we introduce a COT prompt with crossover and mutation operations aiding
LLMs in 
developing improved optimized code.

To evaluate the effectiveness of \tool, we conduct experiments on a widely-used benchmark dataset containing both Python and C++ code. We compare \tool against
four representative prompting methods on four popular open-source and closed-source LLMs including  CodeLlama~\cite{DBLP:journals/corr/abs-2308-12950}, Gemini~\cite{Gemini}, ChatGPT~\cite{ChatGPT}, and GPT-4~\cite{GPT4}. The experimental results demonstrate the effectiveness of \tool in improving code efficiency. 
Our approach achieves a significant boost in program execution efficiency, surpassing the performance of all baseline methods by 8.75\% $\sim$ 28.06\% and 1.15\% $\sim$ 9.56\% with different LLMs in terms of top-5 speedup rate metric on Python and C++, respectively, 


We summarize our contributions as follows.
\begin{enumerate}
    \item To the best of our knowledge, we are the first to explore the code optimization task from a search perspective and propose to enhance LLMs with search-based methods for the task.
    \item We propose a novel framework \tool for effectively guiding LLMs towards identifying efficient optimization methods in the vast search space, by integrating
    execution-based representative sample selection, adaptive optimization pattern retrieval, and genetic operator-inspired chain-of-though prompt.
    \item Extensive experiments demonstrate the effectiveness of \tool in improving code efficiency compared with baseline methods across different LLMs. 
\end{enumerate}
\section{Proposed Framework}\label{sec:med}

\subsection{Overview}
Fig.~\ref{fig:overview} presents the overview of the proposed framework \tool. 
\tool follows the evolutionary search paradigm that first generates
initial solutions and then iteratively selects the fittest candidates while breeding new ones until termination criteria are met.

To start, \tool acquires the initial seed optimized code 
of the given slow code $S_t$ using existing optimization techniques. 
1) In the execution-based representative sample selection part, \tool evaluates the fitness
of
current optimized code, and selects the representative samples that contain distinct and effective optimization methods by 
a re-ranking
mechanism.
2) In the adaptive optimization pattern retrieval part, \tool retrieves code optimization patterns from the pattern base based on both the slow code and the selected representative samples, aiming to guide LLMs towards rectifying and progressively enhancing their optimization methods.
3) In the genetic operator-inspired chain-of-thought prompting part, \tool constructs a prompt that leverages crossover and mutation operations to facilitate
LLMs in combining existing optimization methods and developing improved optimized code for $S_t$.
The above procedure can be conducted multiple iterations
until it no longer yields further optimization in the code efficiency or reaches the maximum iteration.


\begin{algorithm}[!tpb]
    \SetAlgoLined
    \footnotesize
    \SetKwInOut{Input}{Input}
    \SetKwInOut{Output}{Output}
    \SetKwInOut{Initialize}{Initialize}
    \SetKwProg{Fn}{Function}{:}{}
    \Input{The slow code need to optimize $s_t$, existing predictions with their execution information $E$, the number of selected representative samples $N_s$, the training data $T$,  }
    \Output{Selected representative samples $RS$, retrieved patterns $P$}
    \Initialize{Initialize  three lists $correct\_list, \ incorrect\_list$}
    \Fn{Sample selection and pattern retrieval}{
    
        \tcp{Execution-based Representative Sample Selection} 
        $E$ = $sort$($E$, key=speedup rate, order=descend) 
        
        \For{all $e \in E$}
        {
            \uIf{$e$.acc $== 1$ and Abstract($e$.code) not in $correct\_list$}
            {
                $correct\_list$.append($e$.code))
                
            }
            \ElseIf{$e$.acc $< 1$}
            {
                $incorrect\_list$.append($e$.code)
                
                $e$.dis= 0
            }
        }
        \If{$len(correct\_list) < N_s$}
        {
            \For{all $e_a \in incorrect\_list$}
            {
                \For{all $e_b \in incorrect\_list$}
                {
                   $e_a$.dis += EditDistance(Abstract($e_a$), Abstract($e_b$)) 
                
                }
            }
            $incorrect\_list$ = $sort$($incorrect\_list$, key=dis, order=ascend)
        }
        $RS$ = ($correct\_list$ + $incorrect\_list$)[:$N_s$]
        
        \tcp{Adaptive Optimization Pattern Retrieval} 
        
        $input\_score$ = BM25(Abstract($s_t$), $T.s_a$)
        
        $sim\_score$, $dif\_score$ = $[0, 0, 0, ..., 0]$, $[0, 0, 0, ..., 0]$
        
        \For{all $e \in RS$}
        {
            $d_s, d_f$ = GetDiff(Abstract($s_t$), Abstract($e$.code)) 
            
            $score\_opt$ = $\frac{1}{2}$ (BM25($d_s$, $T.d_s$)+BM25($d_f$, $T.d_f$))
            
            $sim\_score$ += $score\_opt$
            
            $dif\_score$ += max($score\_opt$)-$score\_opt$
            
        }
        $sim\_temp$ = $\mathop{\arg\max}(sim\_score+input\_score)$
        
        $dif\_temp$ = $\mathop{\arg\max}(dif\_score+input\_score)$
        
        $P$ = ($sim\_temp$, $dif\_temp$)
        
    }
    \Return {$RS$, $P$}
\caption{Sample Selection and Pattern Retrieval}
\label{algo:mygo}
\vspace{-0.1cm}
\end{algorithm}

\subsection{Execution-based Representative Sample Selection}\label{subsec:select}
To enable LLMs iteratively refine the optimized code, 
we propose to evaluate the fitness of each sample and provide the selected representative ones to LLMs. As shown in Fig.~\ref{fig:overview}, the execution-based representative sample selection module mainly contains two steps, including (a) fitness estimation and (b) fitness-based re-ranking.

\textbf{Fitness Estimation.}
In order to focus LLMs'
search directions towards the most efficient optimization method, we propose to quantitatively assesses the fitness of each sample based on its accuracy and speedup rate. 
\tool
evaluates current optimized code on a set of \textit{public} test cases. 
In alignment with the previous work~\cite{li2022competition,DBLP:conf/iclr/ZhangCSDTG23}, 
two separate test cases are used in the prediction and evaluation process, respectively, including \textit{public} test cases and \textit{private} test cases.
This setup avoids the leakage of the test case information.

\textbf{Fitness-based Re-ranking.}
Based on the collected fitness information, \tool first prioritizes samples that are both correct and have a high speedup rate.
Specifically, as shown in Algorithm~\ref{algo:mygo},  \tool sorts all the
code snippets based on their speedup rate and divides them into correct and incorrect groups according to the
accuracy (Line 2). To consider the combinations of optimization patterns, we then propose to 
select distinct samples from the correct group as representative samples.
As shown in Line 4 to Line 7, \tool abstracts the correct code
based on the ASTs (Abstract Syntax Trees), and ensures that only one sample with identical abstractions can be chosen.
Considering that incorrect code can also provide hints to LLMs for avoiding the same errors, we involve incorrect code as representative samples. 
Specifically, \tool calculates the edit distance between the abstract code, and prioritizes the incorrect code with distance sum.
The prioritized incorrect code tails
the selected correct code.
The top $N_s$ samples are retained as the selected representative samples $RS$, while the remaining samples are discarded. 

\subsection{Adaptive Optimization Pattern Retrieval}
The adaptive optimization pattern retrieval module aims at providing
LLMs with effective optimization patterns 
to facilitate the generation of improved optimized code. The retrieved optimization patterns are expected to provide hints to LLMs towards generating correct and more efficient code. We propose to involve both
the input slow code $s_t$ and selected representative samples $RS$ to adaptively retrieve effective optimization patterns for LLMs. 
Specifically, the retrieval part considers both the optimization methods that are semantically similar to $RS$ for rectifying potential errors, and those different from $RS$ for drawing inspiration from unexploited optimization methods. As shown in Fig.~\ref{fig:overview}, the adaptive optimization pattern retrieval module mainly contains two steps, including (a) fine-grained pattern parsing 
and (b) representative sample-based pattern retrieval.

\textbf{Fine-grained Pattern Parsing.} 
To facilitate accurate retrieval of similar and different patterns, \tool first parses each optimization pair in the training dataset and extracts patterns with fine-grained optimization information to construct the pattern base. Each optimization pair consists of a non-optimized code snippet $s$ and its optimized version $f$. To preserve the general optimization information and eliminate project-specific influences, \tool parses them into ASTs and obtains their abstracted code $s_a$ and $f_a$. Then it identifies the fine-grained optimization information with the ``difflib'' package~\cite{diff} by isolating the abstracted deleted statements $d_s$ from $s_a$, as well as the abstracted added statements $d_f$ from $f_a$. Based on the above process, we obtain the pattern base with fine-grained optimization information including $s_a$, $f_a$, $d_s$, and $d_f$, which is then used to assist the retrieval process.

\begin{figure}
    \centering
    \begin{subfigure}[b]{0.43\textwidth}
        \centering
        \includegraphics[width=1\textwidth]{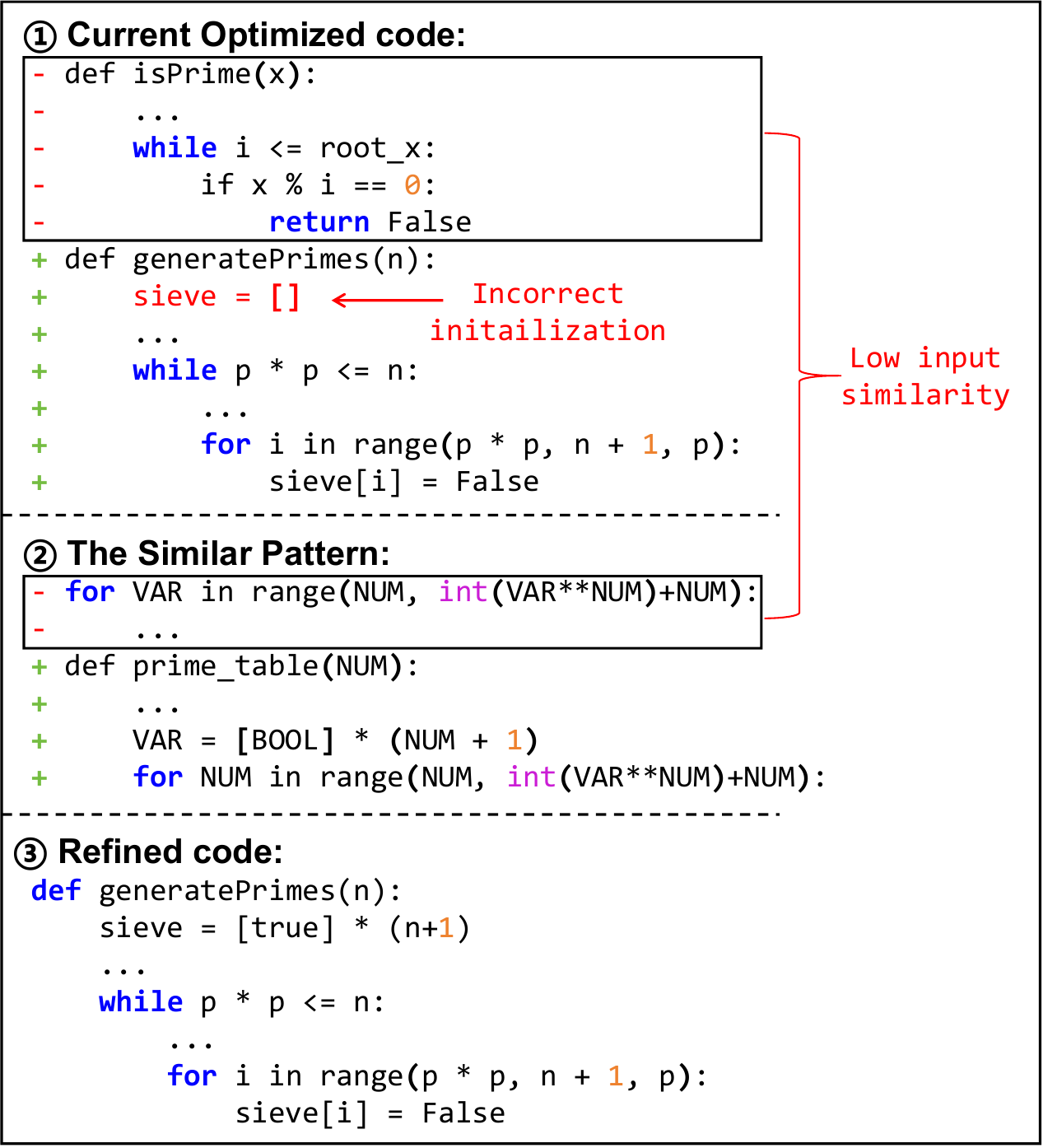}
        \caption{A Python example showing how the similar pattern helps rectify errors in current optimized code.}
    \end{subfigure}
    \hfill
    \begin{subfigure}[b]{0.42\textwidth}
        \centering
        \includegraphics[width=1\textwidth]{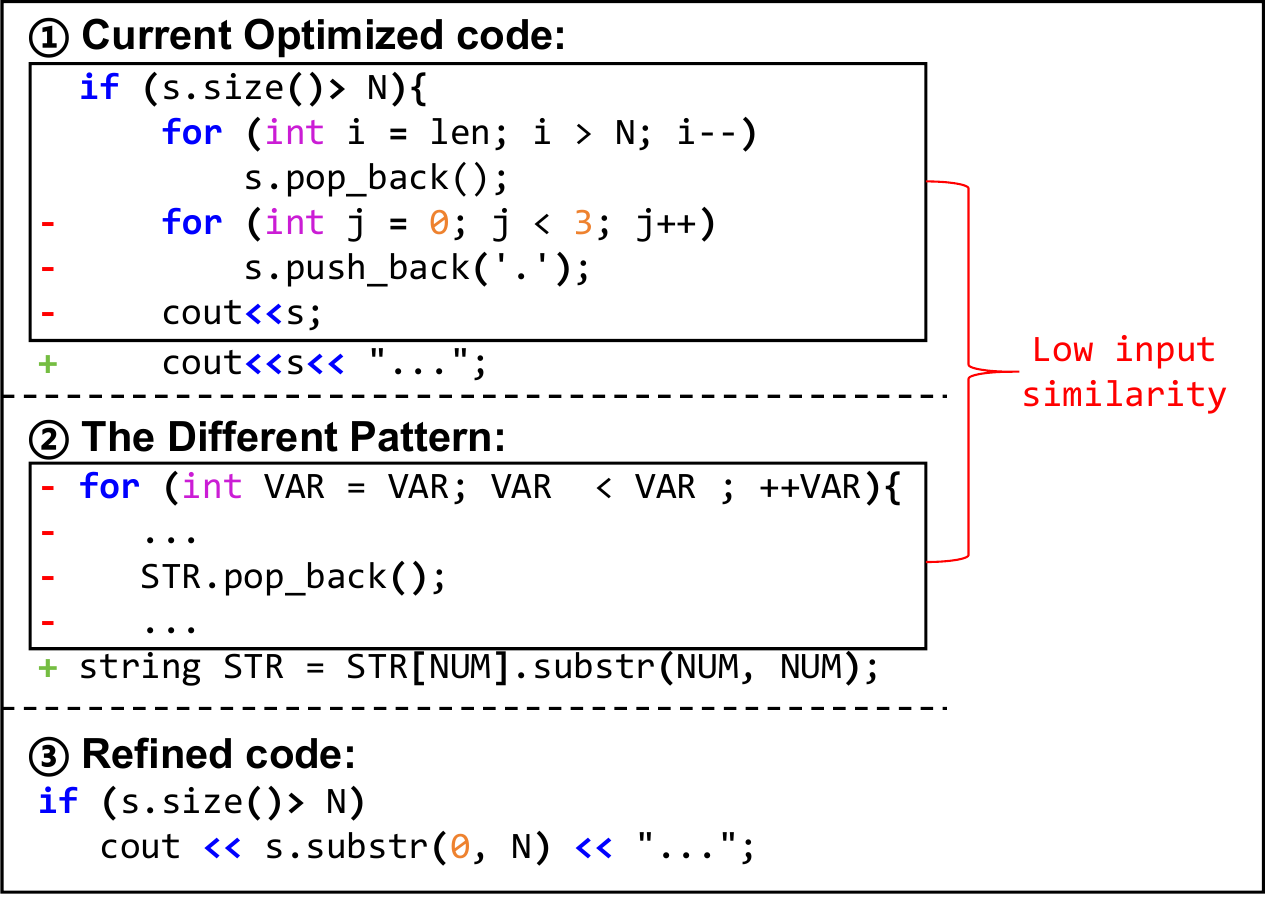}
        \caption{A C++ example showing how the different pattern helps find the unexploited optimization method. 
        }
    \end{subfigure}
    \caption{Examples for illustrating the two kinds of retrieved patterns in the adaptive optimization pattern retrieval module.}
    \label{fig:example}
    \vspace{-0.4cm}
\end{figure}

\textbf{Representative Sample-based Pattern Retrieval.}
The representative sample-based pattern retrieval method aims at guiding LLMs in refining the current optimized code $RS$ in two ways.
Specifically, as shown in Fig.~\ref{fig:example}, current optimized code may contain incorrect and insufficient optimizations. To address these issues, we propose to adaptively retrieve separate
patterns that are semantically similar to and different from $RS$, respectively.

First, 
as shown in Fig.~\ref{fig:example} (a)\circled{1}, the current optimized code intends to use the Eratosthenes Sieve algorithm to reduce the time complexity. However, as LLMs do not learn this optimization method well, the optimized code contains incorrect initialization, 
which will lead to an out-of-index error. The similar pattern presented in Fig.~\ref{fig:example} (a)\circled{2} shows a correct implementation of this algorithm that can provide hints for LLMs to rectify the errors. However, the pattern is hard to be retrieved based solely on its semantic similarity to the input part due to the relatively low similarity degree.
To effectively capture similar patterns,
\tool leverages $RS$ and retrieves the pattern that not only exhibits similar abstracted slow code with $s_a$ but also possesses similar optimized parts (i.e., abstracted deleted statements $d_s$ and abstracted added statements $d_f$) to those in $RS$. Specifically, as shown in Algorithm~\ref{algo:mygo}, \tool first measures the input similarity $input\_score$ between the abstracted code of $s_t$ and the abstracted code snippets $s_a$ in the pattern base by the BM25~\cite{robertson2009probabilistic} (Lines 20). Subsequently, for each representative sample in $RS$, \tool extracts its abstracted deleted statements $d_s$ and abstracted added statements $d_f$, and calculates the similarity score $score\_opt$ of the optimized part with each pattern in the pattern base (Lines 22-24). Then the score is added to the $sim\_score$ (Line 26). Finally, \tool integrates $sim\_score$ with $input\_score$ and the pattern with the highest score is selected as a similar pattern (Line 28).

Second, 
as illustrated in Fig.~\ref{fig:example} (b)\circled{1}, the current optimized code presents
insufficient optimization as it only focuses on the optimization of the second ``for'' loop while overlooking the optimization of the ``pop\_back()'' statement. Correspondingly, the different pattern in Fig.~\ref{fig:example} (b)\circled{2} presents an effective optimization method by employing the ``substr()'' API. 
However, this pattern displays low semantic similarity to the input part, and is hard to be retrieved.
To this end, \tool retrieves the pattern that exhibits similar abstracted slow code $s_a$ but employs different optimization methods compared to those present in $RS$. 
Similarly, \tool also first calculates the similarity score $input\_score$ of the input part and $score\_opt$ of the optimized part based on $RS$.
Then the value of $score\_opt$ is inverted and added to $dif\_score$  to prioritize patterns exhibiting lower similarity of the optimized part (Line 26). Ultimately, the pattern with the highest score of the sum of $dif\_score$ and $input\_score$ is selected as the different pattern (Line 29).
These two retrieved patterns $P$ are then integrated into the prompt to guide the generation of new optimized code. 

\begin{figure}
    \centering
    \includegraphics[width=0.43 \textwidth]{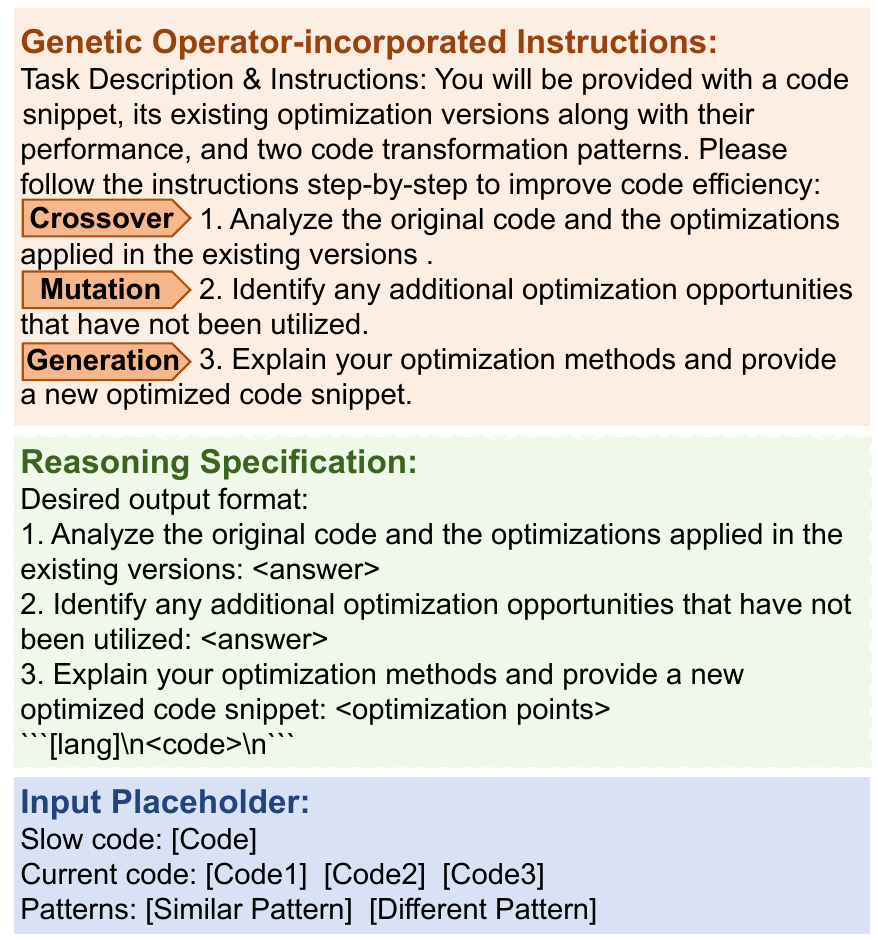}
    \caption{The illustration of the GO-COT prompt. Contents in ``[]'' will be substituted by the corresponding data. The complete prompt can be found in our GitHub repository~\cite{SBLLM}.}
    \label{fig:prompt}
    \vspace{-0.4cm}
\end{figure}

\subsection{Genetic Operator-inspired Chain-of-thought Prompting}
The part aims at guiding
LLMs in integrating different optimization methods from representative samples $RS$ 
and retrieved patterns $P$, and subsequently generating refined
optimized code. Specifically, we propose to aid 
LLMs 
with the evolutionary algorithm's generic operators, and introduce the genetic operator-inspired chain-of-thought (GO-COT) prompt. 
Genetic operators~\cite{sivanandam2008genetic} are inspired by biological evolution principles, and comprise crossover and mutation to synthesize new solutions. These operators facilitate the combination of advantageous traits, promote exploration, and generate superior solutions. 
Specifically, 
we construct the GO-COT prompt, as depicted in Fig.~\ref{fig:prompt}. The prompt 
consists of three main components, including 
genetic operator-incorporated instructions, reasoning specification, and input placeholder.

The \textbf{Genetic Operator-incorporated Instructions} component illustrates the task requirement and outlines the instructions to be executed by LLMs. LLMs are instructed to follow three sequential steps to generate a new code snippet. The first two steps involve combining the advantages observed in the selected representative samples and referring to the retrieved patterns to identify unexploited optimization methods, which correspond to the crossover and mutation operators in the evolutionary algorithm, respectively. Based on these two genetic operators, in the third step, LLMs are required to conclude the optimization methods and generate a new optimized code.
The \textbf{Reasoning Specification} component aims to standardize the format of the output content. LLMs are instructed to follow the given reasoning format step by step and produce the result accordingly.
The \textbf{Input Placeholder} includes the code that LLMs need to optimize along with the representative samples $RS$ and the retrieved patterns $P$. 

With the instructions and provided information in the prompt, LLMs can learn to follow the reasoning strategy and generate a new optimized code step by step. 

\subsection{Evolutionary Optimization}
Based on the aforementioned processes, \tool iteratively generates improved optimized code. As shown in Algorithm~\ref{algo:evol}, at the end of each iteration, the newly generated code will then be integrated with the representative samples in the prompt for the next iteration generation. The iterative refinement process continues until it reaches the convergence condition.
Specifically, if the code has been correctly optimized and the representative samples $RS$ remain unchanged with those in the last iteration, we terminate the process. This is based on the intuition that the prompt in this iteration will be the same as the one in the last iteration and LLMs are less likely to generate better code. 

\resetalgocf
\begin{algorithm}[!tpb]
    \SetAlgoLined
    \footnotesize
    \SetKwInOut{Input}{Input}
    \SetKwInOut{Output}{Output}
    \SetKwInOut{Initialize}{Initialize}
    \SetKwProg{Fn}{Function}{:}{}
    \Input{The slow code need to optimize $s_t$, maximum iteration number $I$}
    \Output{Re-ranked optimized code $Sol$}
    \Fn{Evolutionary optimization}{
        Obtain initialization solutions $Sol$ for $s_t$
        
        \For{all $i \in [1 ... I]$}
        {
            Select representation samples $RS_i$ from $Sol$ and retrieve patterns

            \If{$RS_i == RS_{i-1}$ and $RS_i$ \text{have correctly optimized} $s_t$}
            {
                Break
                
            }   

            Generate new code $NC$ 

            $Sol \gets RS_i \bigcup NC$ 
        }

        Re-rank $Sol$ using the selection part in Algorithm~\ref{algo:mygo}
    }
    \Return {$Sol$}
\caption{The evolutionary optimization process}
\label{algo:evol}
\vspace{-0.1cm}
\end{algorithm}

\section{EXPERIMENTAL setup}\label{sec:setup}

\subsection{Research Questions}
In the evaluation, we focus on the following four research questions:

\begin{enumerate}[label=\bfseries RQ\arabic*:,leftmargin=.5in]
    \item How effective is \tool in improving code efficiency?
    \item What is the fine-grained performance of code generated by \tool across different optimization levels?
    \item What are the contributions of different modules in \tool?
    \item What is the impact of different hyper-parameters on the performance of \tool?
\end{enumerate}

To study RQ1, we conduct a comprehensive evaluation of \tool by comparing
with four representative baseline methods across four popular LLMs, aiming to provide a thorough assessment across language models with different parameter sizes and capabilities. For RQ2, we delve into the analysis of the proportion of generated code across different accuracy and speedup rate levels, including cases where the code is not correct, correct but not faster than slow code, faster than slow code but not faster than human reference and faster than human reference. Additionally, we investigate how much code generated by \tool could be faster than reference code derived by human developers. 
For RQ3, we remove different parts in \tool to assess their individual contributions. For RQ4, we explore the influence of various hyperparameters by varying the number of representative samples in the prompt and the number of maximum iterations.




\subsection{Datasets} 
In this work, we evaluate \tool on the widely-used PIE~\cite{DBLP:journals/corr/abs-2302-07867} dataset which contains two programming languages (i.e., Python and C++). 
The two popular programming languages
are critical for code optimization evaluation.
Python is a dynamic language known for its slow execution speed~\cite{DBLP:conf/osdi/BergerSP23}. 
In contrast, C++ is a statically typed, compiled language renowned for its high performance, especially when leveraging 
the O3 optimization option. By applying the proposed
optimization, we can validate whether the optimized methods generated by \tool are trivial and can be achieved through compiler optimization techniques. 
The PIE dataset is derived from CodeNET~\cite{DBLP:conf/nips/Puri0JZDZD0CDTB21}, which is curated from an online judge system and contains 3,474 programming problems. 
Here we craft the public and private test cases for each problem by using the input-output examples in the program description in CodeNET and the test cases provided by AlphaCode~\cite{li2022competition}, respectively. On average, for each problem, we obtain 2.8 public test cases to obtain feedback for \tool and 95.9 private test cases to evaluate the correctness and efficiency of generated code. Each entry in PIE contains a triplet $($problem id, slow code, fast code$)$ written by the same programmer. The Python subset of the PIE dataset comprises 36,857 training samples, 1,940 valid samples, and 986 test samples; while the C++ comprises 77,967 training samples, 2,544 valid samples, and 994 test samples. 

\subsection{Baselines}
To provide a comprehensive evaluation, we experiment on four popular LLMs and compare \tool with four representative prompt methods, with details as below. 

For LLMs, we evaluate the performance of \tool on both open-source and closed-source models, including CodeLlama, Gemini, ChatGPT, and GPT-4. CodeLlama~\cite{DBLP:journals/corr/abs-2308-12950} is a family of open-source large-scale code language models developed by Meta. We use the 34B instruct-tuned version (i.e., CodeLlama-34b-Instruct-hf) for experiments.  ChatGPT~\cite{ChatGPT} and GPT-4~\cite{GPT4} are two popular LLMs developed by OpenAI which show versatile abilities across different fields such as code generation. They are closed-source model and we access it through APIs (i.e., gpt-3.5-turbo-0613 and gpt-4-1106-preview). Gemini~\cite{Gemini} is a recent powerful closed-source LLM developed by Google which shows comparable ability with GPT-4. We also access it based on its official API (i.e., gemini-pro).

As for the prompt methods, we follow previous work~\cite{DBLP:journals/corr/abs-2302-07867} and involve four representative methods including instruction prompting, in-context learning (ICL), retrieval-augment generation (RAG), and chain-of-though (COT). Instruction prompting directly prompts LLMs to generate optimized code without providing other information. In-context learning adds some examples (input-output pair) before the query sample to help the model understand this task. Following prior work~\cite{DBLP:conf/kbse/GaoWGWZL23}, we randomly sample four pairs from the training set to create the examples for ICL. As for retrieval-augment generation (RAG) method, instead of random selection, it retrieves different samples from the training set for different query samples. Specifically, we employ BM25 to select the code from the training set with the highest similarity to the query sample. Lastly, in the chain-of-thought (COT) prompt, we follow~\cite{DBLP:journals/corr/abs-2302-07867,DBLP:journals/corr/abs-2306-17077} and employ prompts that instruct the LLM to first explain how to optimize the program before producing the optimized code. We use the same examples as ICL for COT and manually craft the explanations to aid the LLM in reasoning through COT. The detailed prompts for all baseline methods are provided in our replication package~\cite{SBLLM}.

\subsection{Metrics} To evaluate the correctness and efficiency of optimized code, we follow previous work~\cite{DBLP:journals/corr/abs-2302-07867} and measure the following metrics:


\textbf{Percent Optimized (OPT):} OPT denotes the fraction of code in the test set that demonstrate improvement through a given method.  A program must be at least 10\% faster and correct (i.e., pass all test cases) to contribute, i.e., $\frac{T(s)-T(o)}{T(o)} > 10\%$ and $A(o)=1$, where $T(\cdot)$ and $A(\cdot)$ represent the execution time and accuracy and $o$ and $s$ denote the optimized and slow code, respectively.

\textbf{Speedup Rate (SP):} SP measures the improvement in running time. 
We first calculate the speedup rate of each generated code and then report the average results on the whole test set. If a generated code is either incorrect or slower than the original slow code, we assign a speedup of 1.0 to that example, as the worst-case scenario assumes the original program has a speedup of 1.0. Formally, SP is calculated as follows:
\begin{equation}
\text{SP}=\sum_{i=1}^{n} \frac{T(s_i)}{T(o_i)} \text{\ if $A(o_i) = 1 \wedge T(o_i) \le T(s_i)$ else 1} 
\end{equation}
where $n$ is the size of test set. 



\begin{table*}[t]
    \centering
    \caption{The performance of \tool along with the baselines on two programming languages in terms of Top-1,3,5. Under each metric the best performance is marked as \g gray. ``*'' denotes statistical significance in comparison to the base models (i.e., Wilcoxon-test with $p$-value$<0.05$).}
    \scalebox{0.95}{
    \begin{tabular}{clccccccccccccccc}
    \toprule
        \multicolumn{2}{c}{{\multirow{3}{*}{\textbf{Approach}}}} & \multicolumn{7}{c}{\textbf{Python}} & &  \multicolumn{7}{c}{\textbf{C++ (O3)}} \\
        \cmidrule{3-9}
        \cmidrule{11-17}
        & & \multicolumn{3}{c}{\textbf{OPT@k}}  & & \multicolumn{3}{c}{\textbf{SP@k}} &  & \multicolumn{3}{c}{\textbf{OPT@k}}  & & \multicolumn{3}{c}{\textbf{SP@k}} \\
        \cmidrule{3-5}
        \cmidrule{7-9}
        \cmidrule{11-13}
        \cmidrule{15-17}
         & & Top-1 & Top-3 & Top-5 & & Top-1 & Top-3 & Top-5 & & Top-1 & Top-3 & Top-5 & & Top-1 & Top-3 & Top-5 \\
        \midrule
        \multirow{6}*{\tabincell{c}{CodeLlama}} &  Instruction & 2.94 & 4.87 & 7.00 & & 102.50 & 104.66 & 105.71 & & 0.34 & 1.14 & 1.71 & & 100.72 & 101.22 & 102.32\\
        \specialrule{0em}{1pt}{1pt}
        & ICL & 3.25 & 5.17 & 8.42 & & 104.48 & 107.08 & 109.31 & & 0.80 & 1.14 & 1.60 & & 100.62 & 100.97 & 101.23 \\
        \specialrule{0em}{1pt}{1pt}
        & RAG & 4.97 & 8.32 & 11.76 & & 104.11 & 108.20 & 113.07 & & 0.00 & 0.11 & 0.68 & & 100.14 & 100.31 & 101.05\\
        \specialrule{0em}{1pt}{1pt}
        & COT & 4.67 & 7.91 & 11.46 & & 105.29 & 111.20 & 117.95 & & 0.23 & 0.57 & 0.68 & & 100.76 & 101.04 & 101.11\\
        \specialrule{0em}{1pt}{1pt}
        & \tool & \g 7.00* & \g 11.36* & \g 13.89* & & \g 111.60* & \g 119.92* & \g 126.70* & & \g 2.62* & \g 3.08* &  \g 4.44* & & \g 102.76* & \g 102.98* & \g 103.47*\\
        \specialrule{0em}{1pt}{1pt}
        \midrule
        \multirow{6}*{\tabincell{c}{Gemini}} &  Instruction & 8.63 & 13.71 & 16.65 & & 115.39 & 125.12 & 128.18 & & 1.03 & 1.83 & 2.40 & & 103.84 & 105.33 & 105.90\\
        \specialrule{0em}{1pt}{1pt}
        & ICL & 7.51 & 11.05 & 12.58 & & 114.27 & 122.58 & 125.87 & & 1.03 & 1.83 & 2.40 & & 104.61 & 105.20 & 107.16 \\
        \specialrule{0em}{1pt}{1pt}
        & RAG & 7.92 & 12.18 & 14.42 & & 112.55 & 120.95 & 125.27 & & 1.03 & 2.17 &  3.08 & & 101.03 & 103.01 & 103.51\\
        \specialrule{0em}{1pt}{1pt}
        & COT & 12.68 & 17.65 & 20.69 & & 125.69 & 136.86 & 143.90 & & 1.37 & 3.20 &  3.77 & & 102.50 & 107.32 & 108.50\\
        \specialrule{0em}{1pt}{1pt}
        & \tool & \g 26.47* & \g 28.09* & \g 28.40* & & \g 153.99* & \g 157.84* & \g 158.22* & & \g 4.90* & \g 6.39* &  \g 6.73* & & \g 110.78* & \g 111.74* & \g 112.07*\\
        \specialrule{0em}{1pt}{1pt}
        \midrule
        \multirow{6}*{\tabincell{c}{ChatGPT}} &  Instruction & 4.97 & 7.61 & 9.94 & & 104.44 & 106.62 & 109.75 & & 0.91 & 2.28 & 4.79 & & 101.42 & 102.96 & 108.45\\
        \specialrule{0em}{1pt}{1pt}
        & ICL & 9.84 & 13.59 & 17.85 & & 115.39 & 120.53 & 127.40 & & 1.14 & 2.28 & 4.34 & & 104.25 & 105.99 & 110.16 \\
        \specialrule{0em}{1pt}{1pt}
        & RAG & 12.37 & 17.85 & 22.62 & & 116.68 & 121.74 & 128.61 & & 1.14 & 2.40 & 4.34 & & 104.25 & 105.40 & 110.33\\
        \specialrule{0em}{1pt}{1pt}
        & COT & 27.89 & 34.79 & 38.34 & & 147.59 & 158.21 & 164.59 & & 3.08 & 5.71 & 6.62 & & 105.54 & 113.22 & 114.80\\
        \specialrule{0em}{1pt}{1pt}
        & \tool & \g 34.58* & \g 37.32* & \g 38.44 & & \g 175.26* & \g 179.83* & \g 182.08*  & & \g 5.93* & \g 7.75 &  \g 8.21 & & \g 120.84* & \g 121.83* & \g 122.26*\\
        \specialrule{0em}{1pt}{1pt}
        \midrule
        \multirow{6}*{\tabincell{c}{GPT-4}} &  Instruction & 21.43 & 29.59 & 32.65 & & 122.35 & 126.36 & 127.33 & & 4.44 & 8.89 & 11.11 & & 108.53 & 119.02 & 123.69\\
        \specialrule{0em}{1pt}{1pt}
        & ICL & 36.73 & 39.80 & 41.84 & & 173.15 & 175.68 & 181.53 & & 5.56 & 6.67 & 11.11 & & 118.53 & 119.45 & 121.53\\
        \specialrule{0em}{1pt}{1pt}
        & RAG & 31.63 & 34.69 & 36.73 & & 161.83 & 165.64 & 170.51 & & \g 6.67 & 6.67 & 14.44 & & 132.90 & 134.56 & 145.20\\
        \specialrule{0em}{1pt}{1pt}
        & COT & 32.65 & 34.69 & 37.76 & & 171.09 & 176.79 & 178.34 & & 5.56 & 8.89 & 13.33 & & 125.41 & 133.09 & 135.40\\
        \specialrule{0em}{1pt}{1pt}
        & \tool & \g 41.84 & \g 45.92* & \g 47.96* & & \g 201.58* & \g 206.17* & \g 209.59* & & 5.49 & \g 9.89 &  \g 16.48 & & \g 143.49 & \g 150.80 & \g 154.76\\
        \specialrule{0em}{1pt}{1pt}
    \bottomrule
    \end{tabular}}
    \label{tab:main}
\end{table*}

\subsection{Implementation Details}
For the hyperparameters of all LLMs, following the previous work~\cite{DBLP:journals/corr/abs-2302-07867}, we set the temperature to 0.7 for all experiments and generate five results by random sampling. For all baseline methods, the optimized code will be re-ranked based on the output probability predicted by the LLMs.
For \tool, after the whole optimization process, we re-rank the optimized code obtained in the last iteration using the selection method in Algorithm~\ref{algo:mygo}.
As for the hyper-parameters of \tool, we set the number of selected representative samples $N_s$ to three and the maximum iteration number to four. The impact of different numbers is discussed in Section~\ref{subsec:RQ4}. For GPT-4, due to the cost limitation, we randomly sample 10\% data in the test set for experiment. 
Following previous work~\cite{DBLP:journals/corr/abs-2302-07867}, we execute each slow and generated program 25 times, and report the average execution results excluding the first run. For the execution environment, we execute Python programs with Python 3.9.12 and compile all C++ programs with GCC version 9.4.0 and C++17 as well as the O3 optimization flag. 
All experiments are conducted on a Linux server (64-bit Ubuntu 20.04) with one 112-core Intel Xeon Platinum 8276 CPU@ 2.20GHz, four NVIDIA A100-40GB GPUs, and 1TB RAM.

\section{Experimental Results}\label{sec:result}

\subsection{RQ1: Comparison with Baselines}\label{subsec:RQ1}
To assess the effectiveness of \tool in improving code efficiency, we compare \tool with four representative prompt methods on four popular LLMs. Table~\ref{tab:main} presents the performance of \tool along with baseline methods on Python and C++. For each metric, we denote the performance using the Top@k metric, where k represents the number of generated code snippets considered.

\textbf{Comparison of the OPT metric.} 
As shown in Table~\ref{tab:main}, \tool demonstrates considerable improvements over the baseline methods across both languages and all LLMs. For example, when compared to the strongest baseline method, COT, \tool achieves an average improvement of 5.11\% and 2.87\% in OPT@5 on Python and C++, respectively. These results demonstrate the effectiveness of \tool in identifying efficient optimization methods within the vast search space. Besides, by comparing the improvements across different LLMs, we observe that \tool achieves higher enhancements on more powerful LLMs such as ChatGPT and GPT-4. For instance, on Python, \tool enhances the OPT@1 of CodeLlama and GPT-4 by 2.37\% and 9.19\%, respectively. This discrepancy can be attributed to the limited instruction-following capability
and context size of CodeLlama, which might make it not fully comprehend the instructions and input information provided by \tool. In contrast, powerful LLMs such as
GPT-4 exhibit better understanding of the GO-COT prompt, 
enabling them to generate superior optimization methods step-by-step.

\begin{figure}[t]
    \centering
    \begin{subfigure}[h]{0.24\textwidth}
        \centering
    	\includegraphics[width=1 \textwidth]{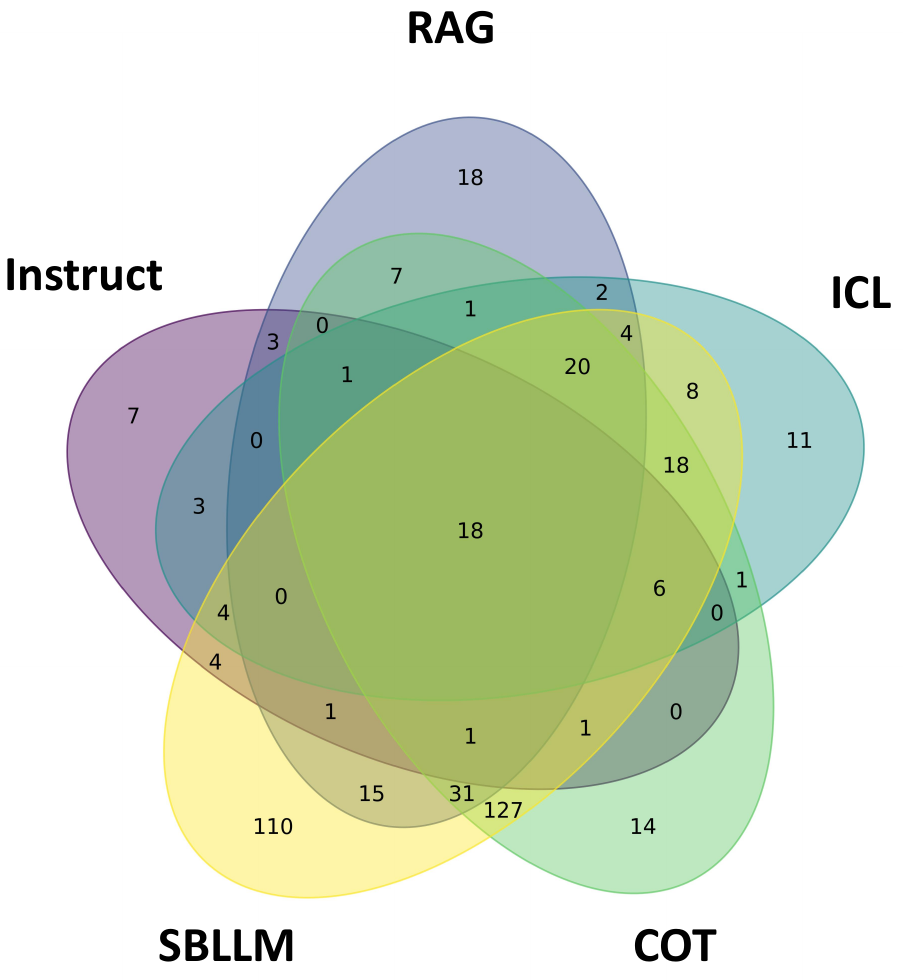}
    	\caption{Python.}
    \end{subfigure}
    \hfill
    \begin{subfigure}[h]{0.24\textwidth}
        \centering
        \includegraphics[width=1 \textwidth]{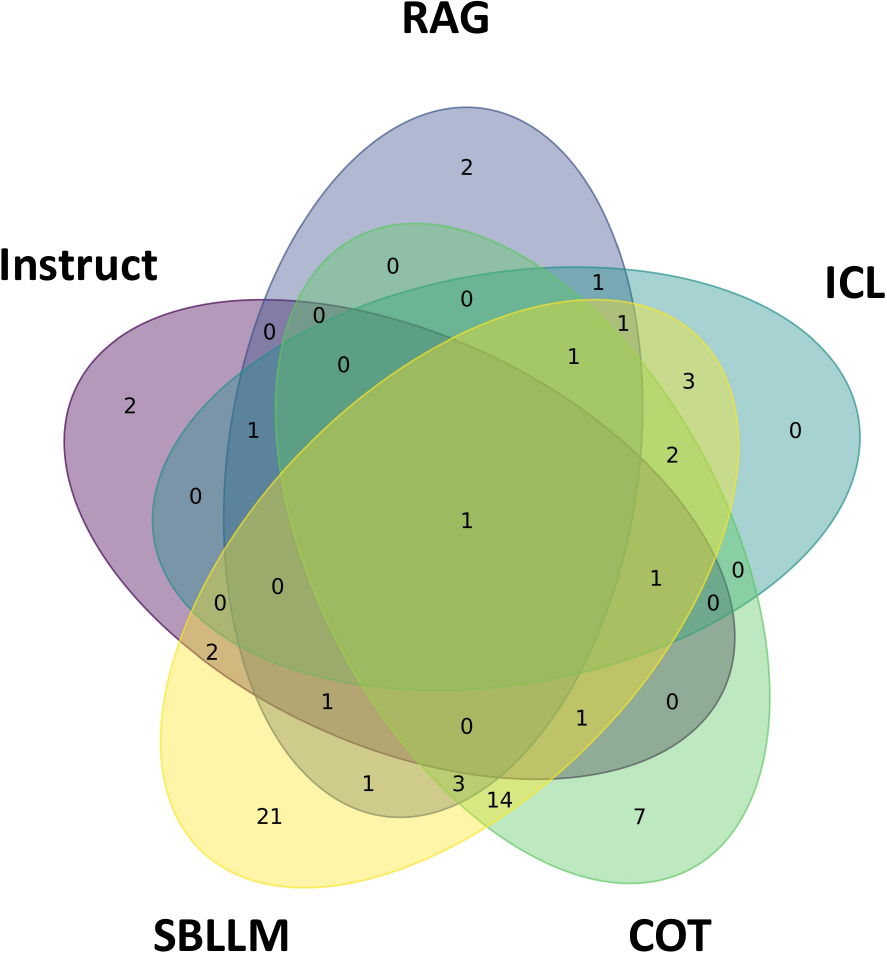}
        \caption{C++ (O3).}
    \end{subfigure}
    \caption{Venn diagram of optimized code provided by \tool and baseline methods on ChatGPT.}
    \label{fig:venn}
\vspace{-0.3cm}
\end{figure}

In addition to measuring the number of optimized code generated by each approach, we also follow previous work in program repair~\cite{DBLP:conf/sigsoft/ZhuSXZY0Z21,DBLP:journals/corr/abs-2210-14179,DBLP:conf/icse/PengGGHL24} and investigate the overlap of optimized code among different methods. 
Fig.~\ref{fig:venn} illustrates the unique optimized code of the top-1 predictions achieved by \tool and four baseline methods on ChatGPT, represented through Venn diagrams. 
Notably, \tool identifies 110 and 21 unique optimized code in Python and C++, respectively, while the other approaches only yield $7 \sim 18$ and $0 \sim 7$ unique optimized code across the two languages. This indicates that the contribution of the iterative refinement process cannot be replaced by the combination of existing prompting approaches. 

\textbf{Comparison of the SP metric.} 
As for the speedup rate, by analyzing the top-5 prediction results in Table~\ref{tab:main}, we observe that \tool can boost program execution efficiency by up to 209.59\% and 154.76\%, outperforming all baselines by 8.72\% $\sim$ 28.06\% and 1.15\% $\sim$ 9.56\% on Python and C++, respectively. The improvement of \tool over baselines is even more significant for top-1 predictions where \tool surpasses COT on GPT-4 by 30.49\% and 18.08\% on two languages, respectively, which can be attributed to our execution-based representative selection method 
in \tool. These results demonstrate that \tool can effectively guide the LLMs progressively moving towards the better optimization method and generating more efficient code.

\begin{figure*}[t]
    \centering
    \begin{subfigure}[h]{0.49\textwidth}
        \centering
    	\includegraphics[width=1 \textwidth]{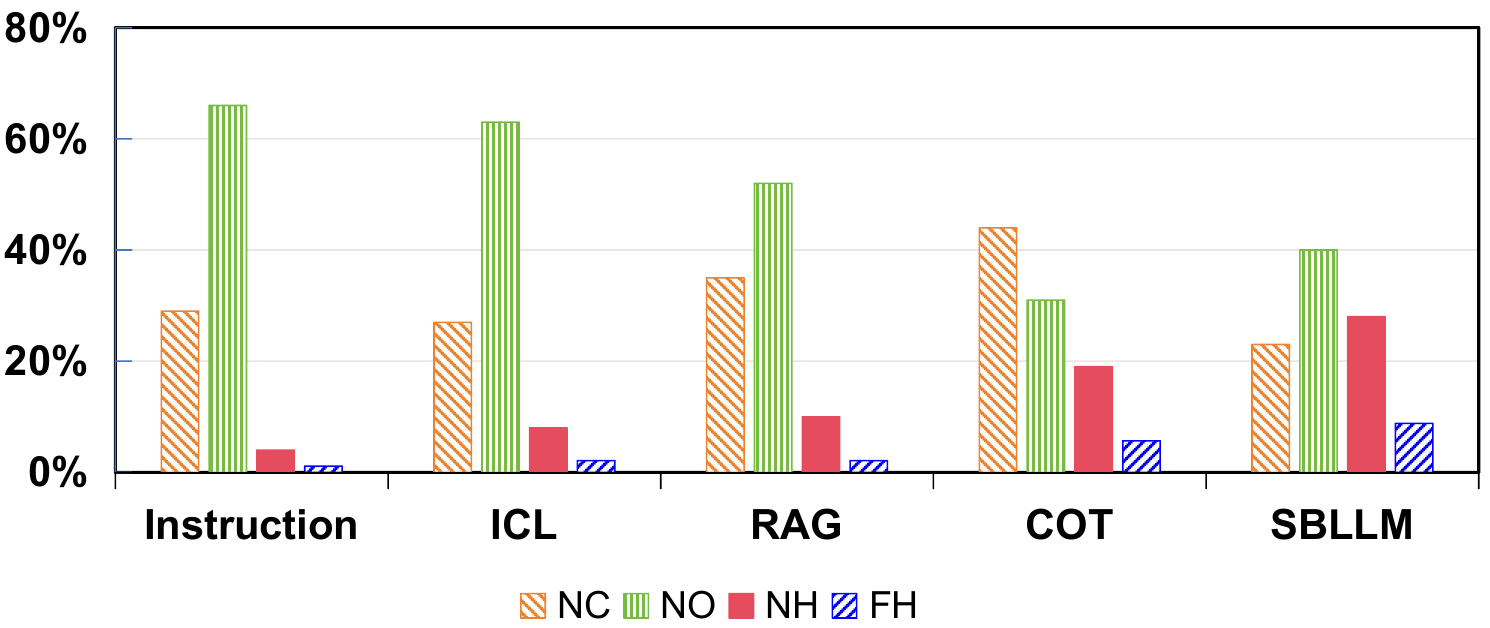}
    	\caption{The proportion of different optimization level on Python.}
    \end{subfigure}
    \hfill
    \begin{subfigure}[h]{0.49\textwidth}
        \centering
        \includegraphics[width=1 \textwidth]{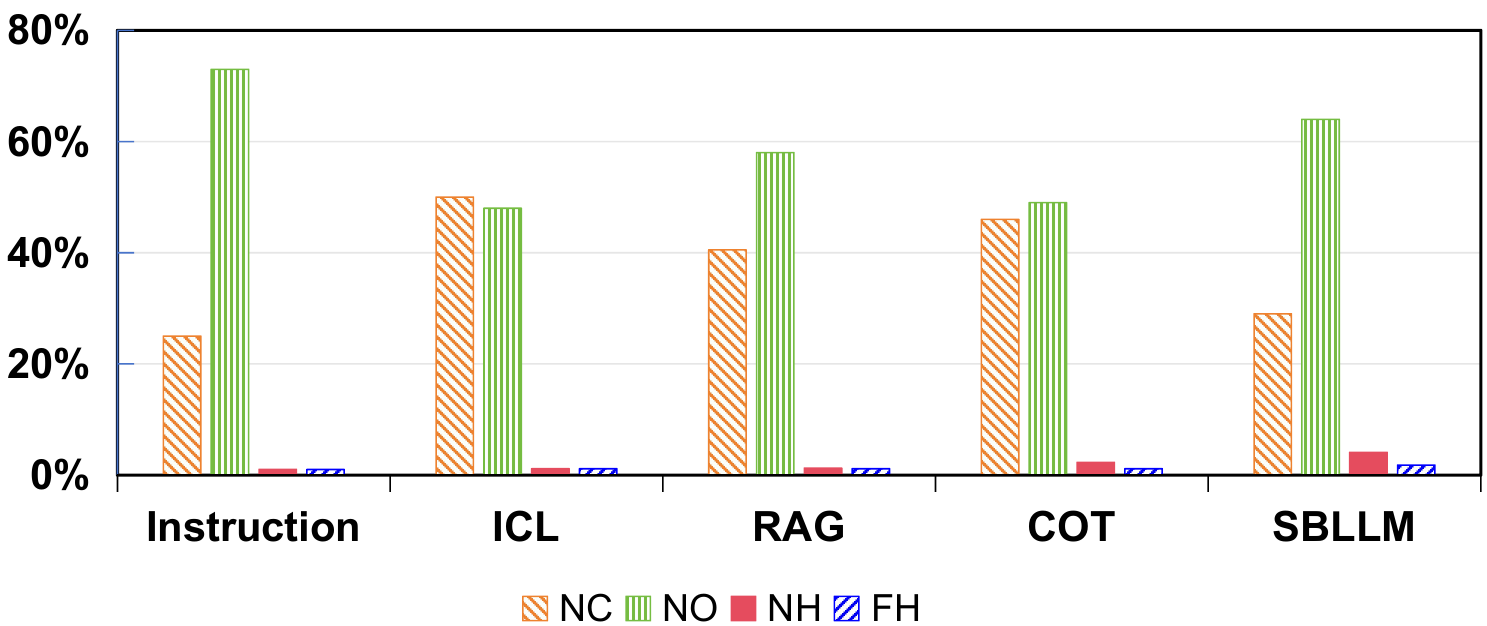}
        \caption{The proportion of different optimization level on C++ (O3).}
    \end{subfigure}
    \caption{The proportion of top-1 prediction based on ChatGPT with different optimization levels on Python and C++, respectively. ``NC'', ``NO'', ``LH'', and ``FH'' indicate the code is not correct, correct but not optimized, optimized but lower than human reference and faster than human reference, respectively.}
    \label{fig:proportion}
    \vspace{-0.4cm}
\end{figure*}

\begin{tcolorbox}[width=\linewidth,boxrule=0pt,top=1pt, bottom=1pt, left=1pt,right=1pt, colback=gray!20,colframe=gray!20]
\textbf{Answer to RQ1:} 
\tool successfully optimizes the most code snippets compared to all baselines across different LLMs. It boosts program execution efficiency by up to 209.59\% and 154.76\%, outperforming the top-5 speedup rate of all baselines by 8.75\% $\sim$ 28.06\% and 1.15\% $\sim$ 9.56\%  on Python and C++, respectively, 
\end{tcolorbox}


\subsection{RQ2: Fine-grained Performance Analysis}\label{subsec:RQ2}
In this RQ, we study the fine-grained proportion of generated code across different levels of accuracy and speedup rates. To achieve this, we classify the generated code snippets into four distinct levels: not correct (NC), correct but not optimized (NO), optimized but not faster than the human reference (NH), and faster than the human reference (FH). Following the OPT definition, we consider code that is at least 10\% faster than the human reference as falling into the FH category. The experimental results are depicted in Fig.~\ref{fig:proportion}. Due to space limitations, we only present the top-1 results of each method on ChatGPT, while the results for other models can be found in our GitHub repository~\cite{SBLLM}. 

\begin{figure}[t]
    \centering
    \begin{subfigure}[h]{0.24\textwidth}
        \centering
    	\includegraphics[width=1 \textwidth]{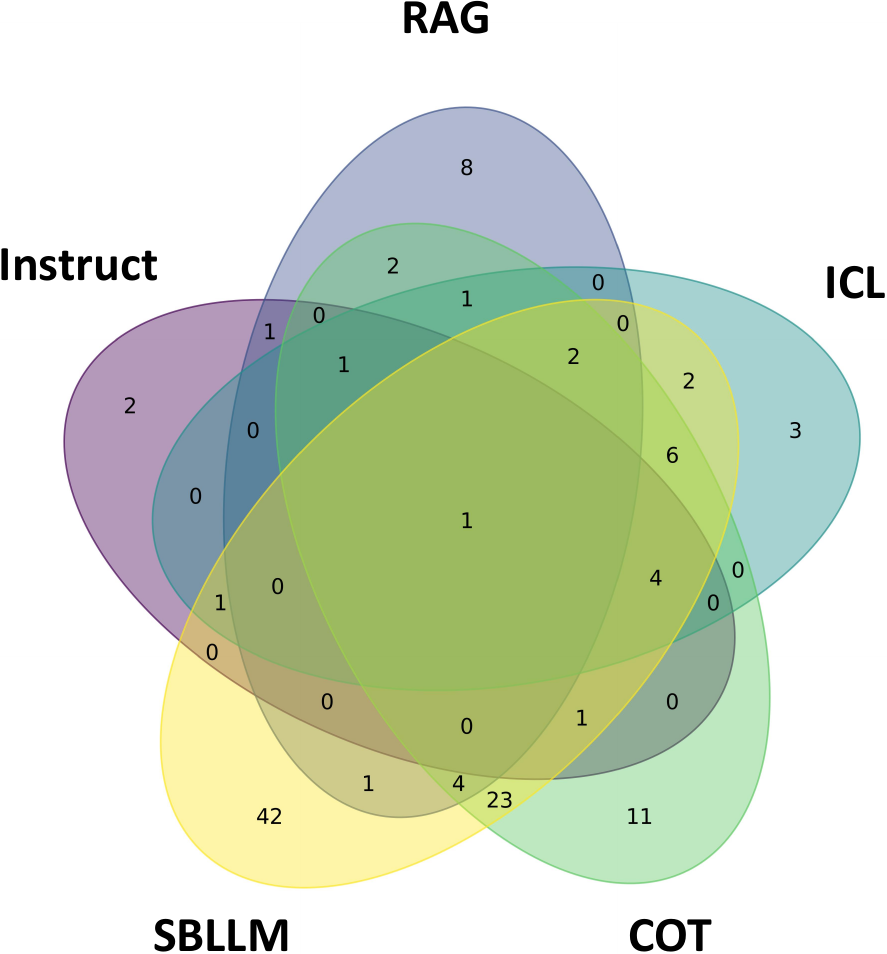}
    	\caption{Python.}
    \end{subfigure}
    \hfill
    \begin{subfigure}[h]{0.24\textwidth}
        \centering
        \includegraphics[width=1 \textwidth]{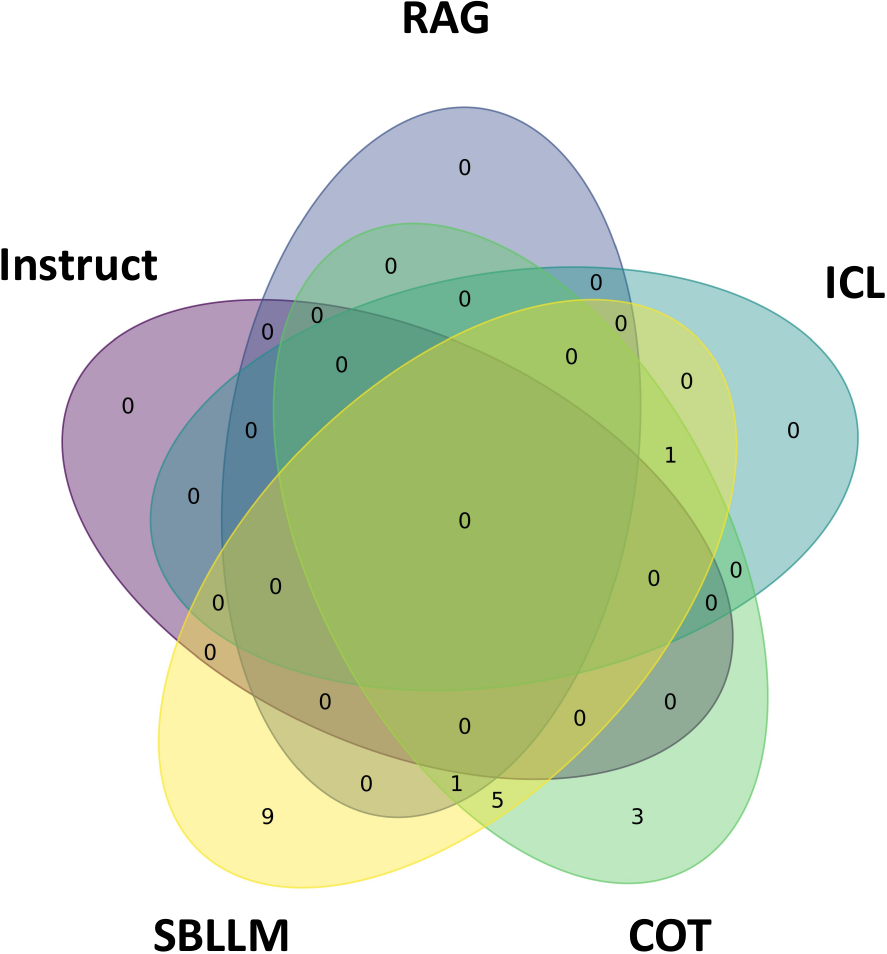}
        \caption{C++ (O3).}
    \end{subfigure}
    \caption{Venn diagram of code faster than human reference provided by \tool and baseline methods on ChatGPT.}
    \label{fig:venn_unique}
    \vspace{-0.45cm}
\end{figure}

As shown in Fig.~\ref{fig:proportion}, we can observe that \tool consistently achieves the relatively lower NC rate and the highest NH and FH rate compared with other methods on both programming languages. 
Specifically, concerning the NC rate, baseline methods generate at least 27.18\% incorrect code for Python. In contrast, SBLLM reduces this rate by 17.18\%. As for the FH rate, 8.82\% and 1.82\% of the code generated by \tool outperforms the human reference on Python and C++, respectively. This represents an obvious improvement over the strongest baseline method named COT, which only achieves FH rates of 5.67\% and 1.14\% on these two programming languages.
Furthermore, we also investigate the overlap of FH code snippets generated by different methods. As shown in Figure~\ref{fig:venn_unique}, we find that \tool achieves 42 and 9 unique improvements on Python and C++, respectively, surpassing the performance of other methods on both languages. 
These findings suggest that the iterative refinement process in \tool facilitates the generation of correct and efficient code solutions.


\begin{tcolorbox}[width=\linewidth,boxrule=0pt,top=1pt, bottom=1pt, left=1pt,right=1pt, colback=gray!20,colframe=gray!20]
\textbf{Answer to RQ2:} 
\tool excels in generating code with the lowest error rate in Python and achieves the highest rate of efficiency surpassing human-written reference across both programming languages.
\end{tcolorbox}

\begin{table*}[t]
    \centering
    \caption{Ablation Study. Under each metric the best performance is marked as \g gray.}
    \scalebox{1}{
    \begin{tabular}{lccccccccccccccc}
    \toprule
        \multicolumn{1}{c}{{\multirow{3}{*}{\textbf{Approach}}}} & \multicolumn{7}{c}{\textbf{Python}} & &  \multicolumn{7}{c}{\textbf{C++ (O3)}} \\
        \cmidrule{2-8}
        \cmidrule{10-16}
        & \multicolumn{3}{c}{\textbf{OPT@k}}  & & \multicolumn{3}{c}{\textbf{SP@k}} &  & \multicolumn{3}{c}{\textbf{OPT@k}}  & & \multicolumn{3}{c}{\textbf{SP@k}} \\
        \cmidrule{2-4}
        \cmidrule{6-8}
        \cmidrule{10-13}
        \cmidrule{14-16}
         & Top-1 & Top-3 & Top-5 & & Top-1 & Top-3 & Top-5 & & Top-1 & Top-3 & Top-5 & & Top-1 & Top-3 & Top-5 \\
        \midrule
         -w/o Selection & 23.65 & 29.14 & 32.39 & & 155.94 & 168.05 & 176.67 & & 3.88 & 4.68 & 5.47 & & 111.25 & 112.68 & 116.09\\
        \specialrule{0em}{1pt}{1pt}
         -w/o Sim Pattern & 32.49 & 35.13 & 36.24 & & 165.59 & 171.49 & 174.01 & & 4.57 &  6.40 & 7.54 & & 117.12 & 120.55 & 121.35\\
        \specialrule{0em}{1pt}{1pt}
         -w/o Dif Pattern & \g  36.04 & 37.16 & 37.97 & & \g 176.79 & 178.85 & 181.19 & &  5.36 & 6.50 & 6.73 & & 115.37 & 116.43 & 117.37\\
        \specialrule{0em}{1pt}{1pt}
         -w/o GO-COT & 34.11 & 36.75 & 36.95 & & 170.18 & 175.29 & 175.71 & & 4.90 & 5.92 & 7.18 & & 112.99 & 114.27 & 116.05 \\
        \specialrule{0em}{1pt}{1pt}
         -w/o ALL & 25.38 & 29.24 & 30.66 & & 153.05 & 161.13 & 164.34 & & 3.64 & 5.81 & 6.73 & & 111.59 & 114.18 & 115.95 \\
        \specialrule{0em}{1pt}{1pt}
         \tool & 34.58 & \g 37.32 & \g 38.44 & & 175.26 & \g 179.83 & \g 182.08  & & \g 5.93 & \g 7.75 &  \g 8.21 & & \g 120.84 & \g 121.83 & \g 122.26\\
        \specialrule{0em}{1pt}{1pt}
    \bottomrule
    \end{tabular}}
    \label{tab:ablation}
\end{table*}

\subsection{RQ3: Ablation Study}\label{subsec:RQ3}
To investigate the individual contribution of different components in \tool, we conduct an ablation study and present the results in Table~\ref{tab:ablation}. Due to the space limitation, we only present the results on ChatGPT, with results for other LLMs presented on our GitHub repository~\cite{SBLLM}.

\textbf{Sample Selection.} To assess the impact of the execution-based representative sample selection component, we replace it with a direct selection method that simply selects samples with the highest speedup rate. As shown in Table~\ref{tab:ablation}, removing our selection strategy results in a noticeable performance decrease. Specifically, the Top-1 OPT drops by 10.93\% and 2.05\% on Python and C++, respectively, which demonstrates the benefits of our sample selection criteria.

\textbf{Pattern Retrieval.} In this experiment, we evaluate the effectiveness of the adaptive optimization pattern retrieval part by removing the two patterns utilized in \tool respectively. When the similar pattern is eliminated, the performance of \tool suffers from an obvious decline, i.e., 9.67\% on the Top-1 speedup rate of Python, which demonstrates the importance of providing LLMs with a similar optimization pattern to refine their optimization methods. Similarly, removing the different patterns leads to a substantial drop of SP metric on C++, as the dissimilarity pattern could assist LLMs in drawing inspiration from unexploited optimization methods and generating more efficient code.

\textbf{GO-COT.} To evaluate the contribution of GO-COT, we remove its crossover and mutation-based genetic instructions and reasoning specification part. Instead, we solely use input placeholder to query LLMs to generate optimized code. The results in Table~\ref{tab:ablation} demonstrate a substantial performance decrease of 5.08\% and 7.85\% on the Top-1 speedup metric for Python and C++, respectively. This indicates that GO-COT is effective in aiding LLMs to combine different optimization methods and generate improved optimized code.

\textbf{All.} In this experiment, we remove all the aforementioned components to evaluate the performance of a direct combination of LLMs and evolutionary search. From table~\ref{tab:ablation}, we can find that removing all of the above components lead to worse results, e.g., a drop of 7.78\% and 1.48\% on the Top-5 OPT on Python and C++, respectively.
This suggests that simply combining LLMs with evolutionary search or increasing the generation number can not yield promising results.


\begin{tcolorbox}[width=\linewidth,boxrule=0pt,top=1pt, bottom=1pt, left=1pt,right=1pt, colback=gray!20,colframe=gray!20]
\textbf{Answer to RQ3:} 
All components in \tool contribute to the performance. Removing the execution-based representative sample selection, adaptive optimization pattern retrieval, or GO-COT leads to substantial performance decreases.
\end{tcolorbox}

\begin{figure}[t]
     \centering
     \begin{subfigure}[h]{0.24\textwidth}
        \centering
    	\includegraphics[width=1 \textwidth]{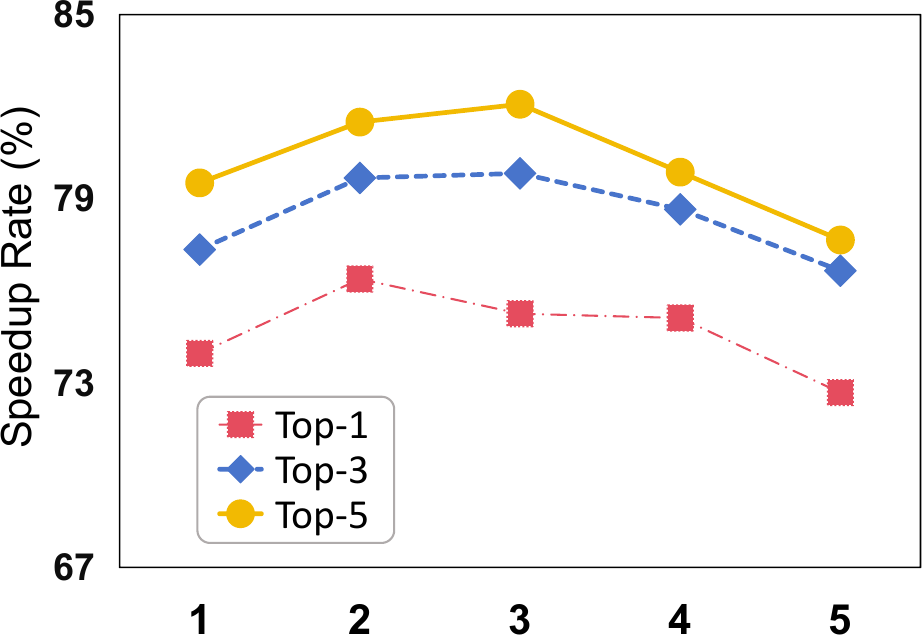}
    	\caption{$N_s$ on Python.}
    \end{subfigure}
    \hfill
    \begin{subfigure}[h]{0.24\textwidth}
        \centering
        \includegraphics[width=1 \textwidth]{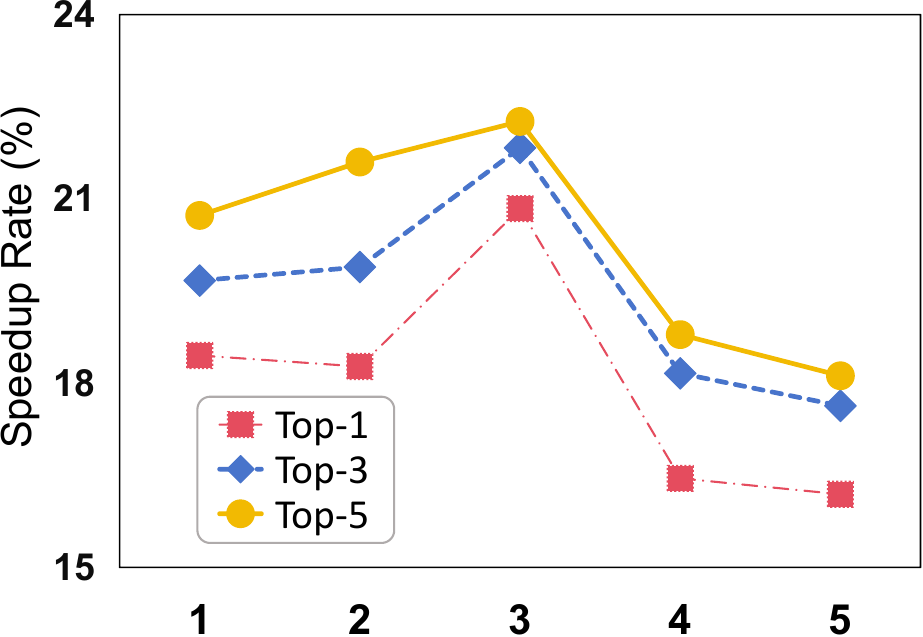}
        \caption{$N_s$ on C++ (O3).}
    \end{subfigure}
    \begin{subfigure}[h]{0.24\textwidth}
    \vspace*{0.15cm}
        \centering
        \includegraphics[width=1 \textwidth]{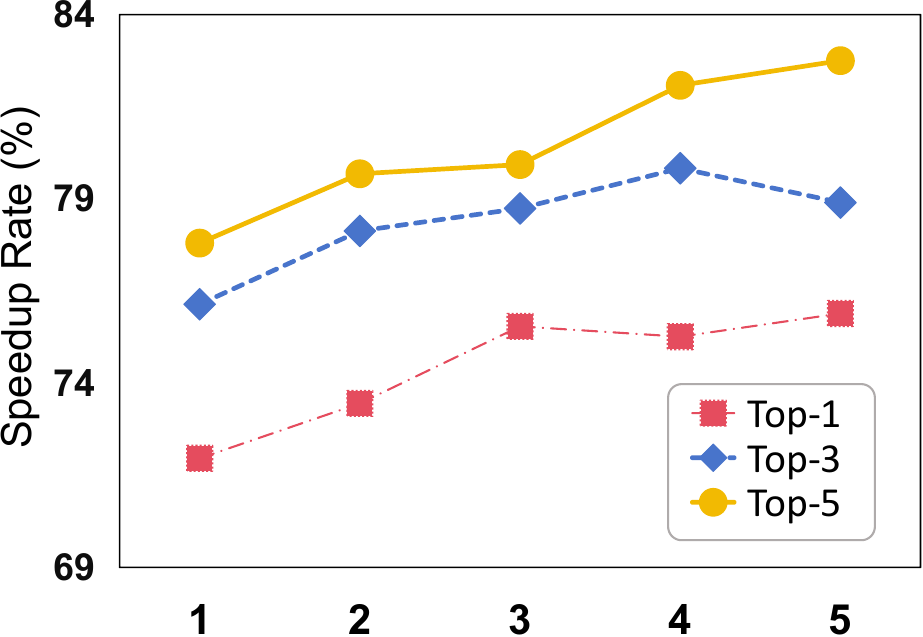}
        \caption{Iteration number on Python.}
    \end{subfigure}
    \hfill
    \begin{subfigure}[h]{0.24\textwidth}
    \vspace*{0.15cm}
        \centering
        \includegraphics[width=1 \textwidth]{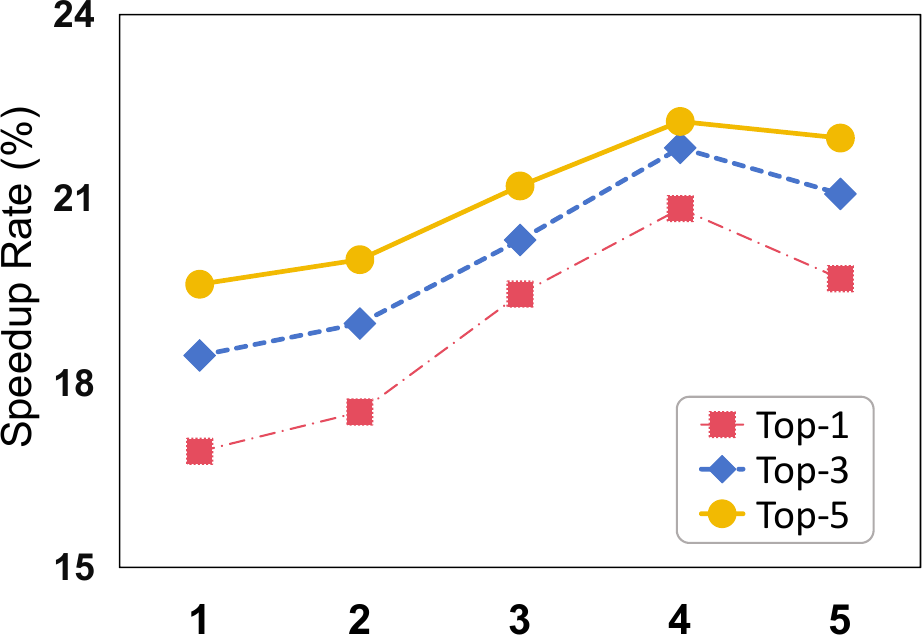}
        \caption{Iteration number on C++ (O3).}
    \end{subfigure}        
    \caption{Parameter analysis.} 
     \label{fig:parameter}
    \vspace{-0.45cm}
\end{figure}

\subsection{RQ4: Parameter Analysis}\label{subsec:RQ4} 
In this section, we investigate the influence of two parameters, namely the number of selected examples ($N_s$) and the maximum number of iterations, on the performance of \tool. Similar to Section~\ref{subsec:RQ3}, the results presented in this section are also based on ChatGPT and the remaining results can be found in the GitHub repository~\cite{SBLLM}.

\textbf{Number of selected representative samples.} We study the effect of $N_s$ by varying it from 1 to 5. As depicted in Fig.~\ref{fig:parameter} (a) and (b) for both Python and C++, \tool exhibits optimal performance when $N_s$ is set to 3. Choosing larger or smaller values does not yield improved results. We suggest that it is because insufficient selected samples may result in inadequate information, while an excessive number of samples may introduce redundancy, which may hinder the model's ability to effectively utilize the provided information~\cite{DBLP:journals/corr/abs-2307-03172}.

\textbf{Number of iterations.} We evaluate the performance of \tool across different iterations by setting the maximum iteration to five. The corresponding results are presented in Fig.~\ref{fig:parameter} (c) and (d). From the results, we can observe that \tool demonstrates increasingly better performance with each iteration on Python; while on C++ it achieves the peak at the fourth iteration since excessively large iterations may introduce the risk of overfiting on public test cases. Consequently, we set the maximum iteration as four. 

\begin{tcolorbox}[width=\linewidth,boxrule=0pt,top=1pt, bottom=1pt, left=1pt,right=1pt, colback=gray!20,colframe=gray!20]
\textbf{Answer to RQ4:} 
\tool achieves the best performance with three representative samples. For the number of iterations, the performance of \tool improves with more iterations initially, but excessively large iteration numbers may cause performance degradation.
\end{tcolorbox}

\section{Discussion}\label{sec:discuss}

\begin{figure}
    \centering
    \includegraphics[width=0.45 \textwidth]{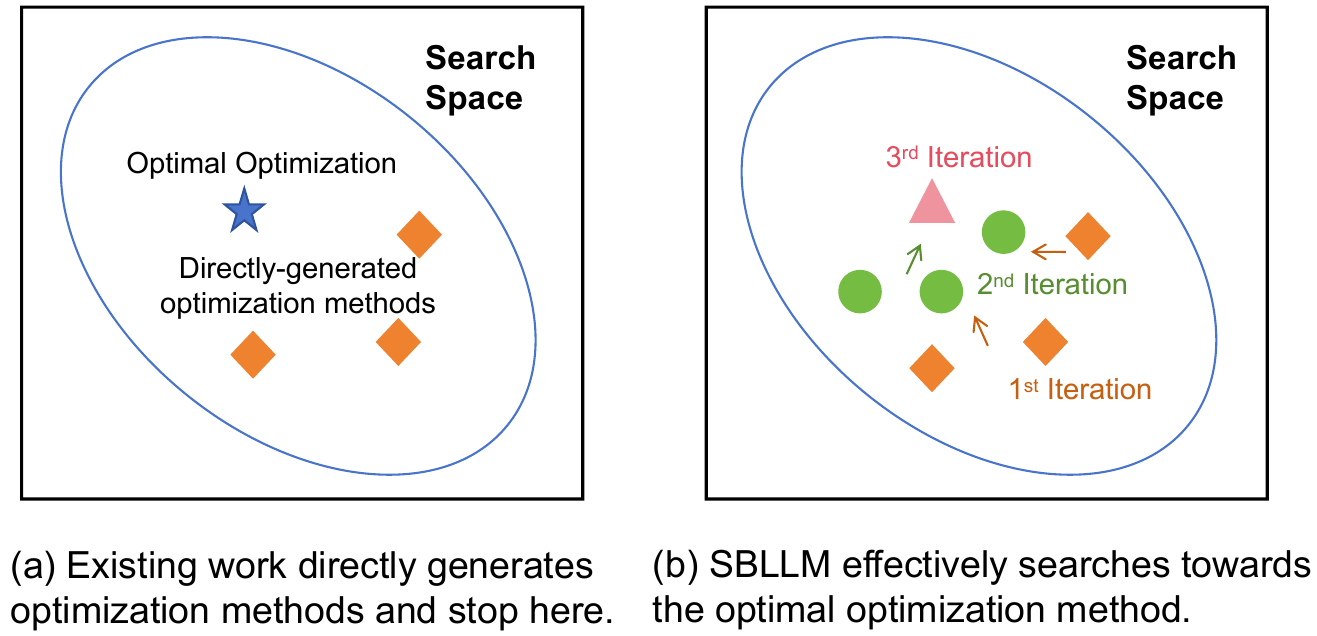}
    \caption{An illustration of the difference of existing work and \tool from the search perspective. 
    }
    \label{fig:search}
    \vspace{-0.45cm}
\end{figure}


\subsection{What makes \tool work?}
\textbf{\tool synergistically integrates evolutionary search into
LLMs.} 
The first advantage of \tool is its novel paradigm which benefits LLMs by incorporating them into the iterative refinement process.
As shown in Fig.~\ref{fig:search} (a), previous LLM-based work directly generates the optimized code. However, due to the complex optimization methods, they are hard to directly yield the optimal solution within such an expansive search space. In contrast, \tool integrates the search-based method which enables selecting effective optimization methods and combining them into improved ones. 
As depicted in Fig.~\ref{fig:search} (b), based on this iterative refinement process, \tool guides LLMs towards the optimal optimization method step-by-step. 

\begin{figure}
    \centering
    \includegraphics[width=0.30 \textwidth]{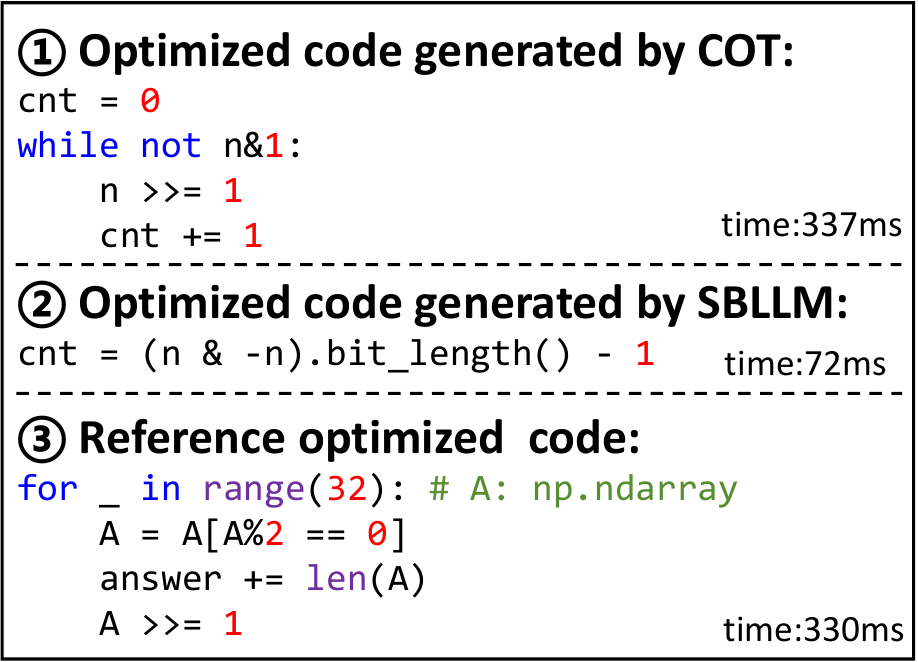}
    \caption{A case showing that \tool can optimize codes with efficient APIs and surpass human reference.}
    \label{fig:case}
    \vspace{-0.45cm}
\end{figure}

\begin{figure}
    \centering
    \includegraphics[width=0.48 \textwidth]{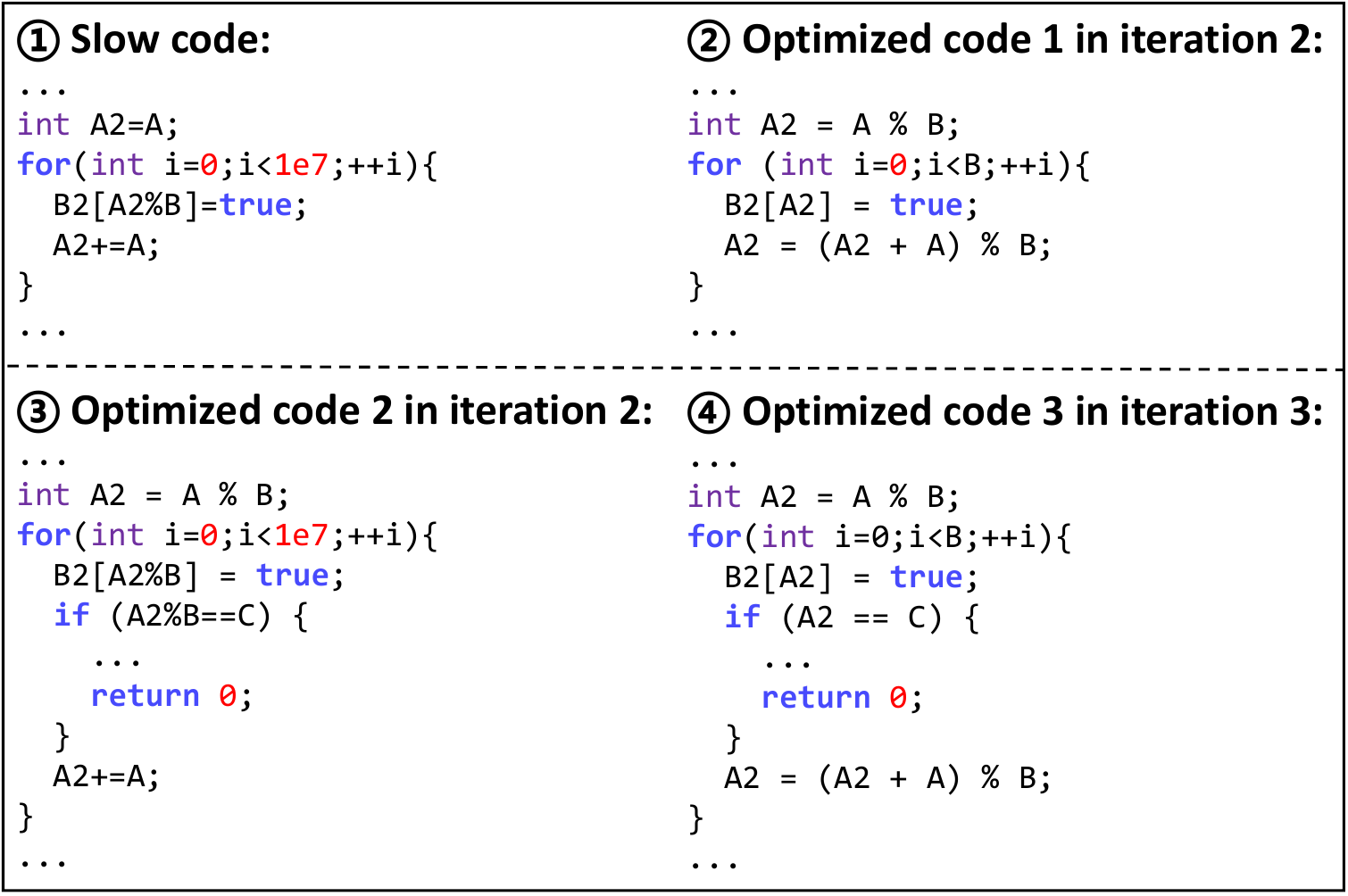}
    \caption{A case showing that LLMs can follow the crossover instruction in GO-COT and combine different optimization methods. The complete case can be found in our GitHub repository.}
    \label{fig:case2}
    \vspace{-0.45cm}
\end{figure}

\begin{figure}
    \centering
    \includegraphics[width=0.35 \textwidth]{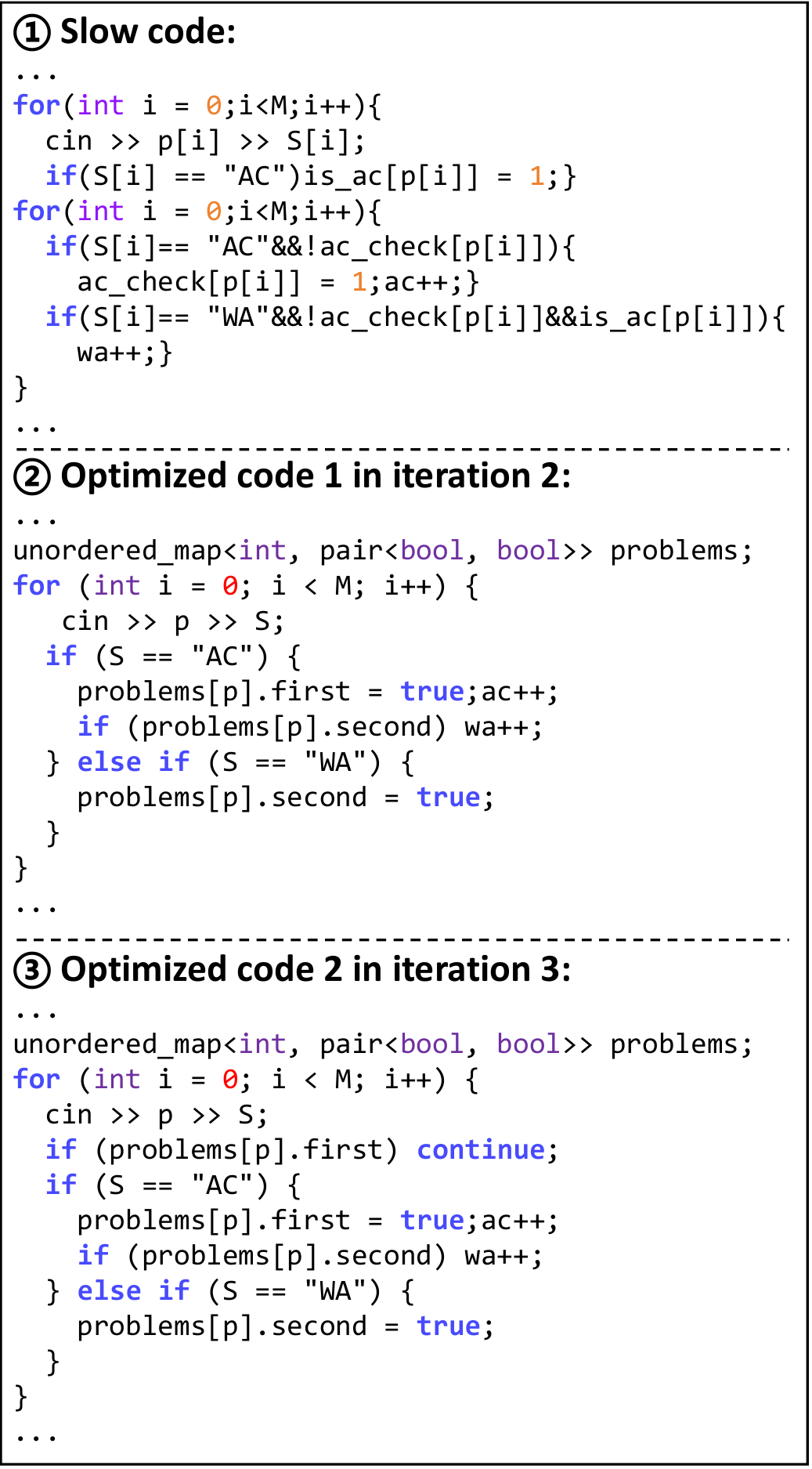}
    \caption{A case showing that GO-COT can follow the mutation instruction in GO-COT and achieve further improvement. The complete case can be found in our GitHub repository.}
    \label{fig:case3}
    \vspace{-0.45cm}
\end{figure}

\textbf{\tool could provide LLMs with effective optimization patterns.} 
The second advantage of \tool lies in its provided effective optimization patterns. LLMs heavily rely on general knowledge acquired during the pre-training phase. Therefore, without explicit external hints, it is hard for them to identify errors in existing optimized code snippets or discover unexploited optimization methods~\cite{DBLP:journals/corr/abs-2310-01798}.
\tool assists LLMs by providing the optimization patterns as hints, thereby guiding them in generating correct and more efficient code.
As shown in Fig.~\ref{fig:case}, the COT prompt with ChatGPT fails to achieve optimal optimization for the slow code due to their limited code optimization knowledge.
\tool achieves better performance by leveraging the retrieved pattern that includes the API ``bit\_length()'', which enables it to even outperform the reference code derived by human developers.

\textbf{\tool could aid LLMs in generating improved optimized code.} Another advantage of \tool is the GO-COT which guides LLMs in effectively leveraging existing optimized code to generate improved ones. As shown in Fig.~\ref{fig:case2}\circled{1}, the original code repeatedly adds A and takes the modulus B.
Although the optimized code snippets generated by LLMs may not directly achieve the optimal optimization method, they may contain various optimization techniques that, when combined, lead to more efficient code. As illustrated in Fig.~\ref{fig:case2}\circled{2} and \circled{3}, these two code snippets improve efficiency by reducing the number of iterations and incorporating an early termination check, respectively. In the next iteration, \tool follows the crossover instruction in the GO-COT prompt and combines them to achieve better performance, as depicted in Fig.~\ref{fig:case2}\circled{4}. Additionally, as shown in Fig.~\ref{fig:case3}\circled{1}, the slow code reads competition problem submissions, tracks accepted (AC) and wrong attempt (WA) statuses using arrays and outputs the counts of AC and WA. In the first optimized version, this code replaces arrays with an $unordered\_map$ to efficiently track AC and WA statuses and updates counts during input processing. Then, as shown in Fig.~\ref{fig:case3}\circled{3}, \tool follows the mutation instruction in the GO-COT prompt and further optimizes the code in the next iteration by skipping the processing for problems that have already been accepted.

\subsection{Threats to Validity}

We identify three main threats to the validity of our study:

1)  \textbf{The selection of languages.} In this paper, we conduct experiments on the PIE dataset containing two widely-used programming languages Python and C++. 
While there are other datasets that contain different programming languages, such as the C\# dataset in DeepDev-PERF~\cite{DBLP:conf/sigsoft/GargMCSW22}, we don't use this dataset since it does not contain test cases. In the future, we will create test cases for this dataset and conduct experiments.

2)  \textbf{The selection of LLMs.} Another threat is the baselines we utilized in our evaluation. We evaluate \tool on four popular and representative LLMs and prompting methods. While there are other existing LLMs~\cite{nijkamp2022codegen,DBLP:journals/corr/abs-2107-03374}, our proposed \tool is model-agnostic and can be easily generalized to different LLMs. Furthermore, our method focuses on how to further boost the initial results directly generated by LLMs.
Therefore, our research is orthogonal to the work that solely generates code such as fine-tuning and one-step prompting methods. 
In future work, we plan to further evaluate the effectiveness of \tool on other LLMs. 

3) \textbf{Potential data leakage.} In this paper, we conduct experiments utilizing the APIs of ChatGPT, GPT-4, and Gemini. However, as these models are closed-source, their training data are not publicly accessible, giving rise to concerns regarding potential data leakage. However, our experiments reveal that directly prompting the LLMs to optimize the code cannot yield promising results. 
Therefore, we believe that the optimization code generated by \tool is not simply from memorizing the training data.


\section{Related work}\label{sec:related}

\subsection{Code Optimization}
Early research in code optimization techniques tends to employ rule-based methods and focus on specific inefficiency types such as software misconfigurations and loop inefficienciess~\cite{krishna2020cadet,DBLP:conf/oopsla/ToffolaPG15,DBLP:conf/msr/NistorJT13,DBLP:conf/kbse/GiavrimisBPB021}. Recently, deep learning-based methods are introduced and achieve promising results. DeepDev-PERF~\cite{DBLP:conf/sigsoft/GargMCSW22} is a pre-trained model that suggests performance improvements in C\# code. 
Chen et al.~\cite{DBLP:journals/corr/abs-2208-05297} introduce a variational auto-encoder that identifies effective code edits for performance. 
RAPGen~\cite{DBLP:journals/corr/abs-2306-17077} leverages OpenAI Codex~\cite{DBLP:journals/corr/abs-2107-03374} to fix C\# code inefficiencies issue in zero-shot. It surpasses DeepDev-PERF in precision without extra training. Supersonic~\cite{DBLP:journals/corr/abs-2309-14846} develops a seq2seq model for code optimization using diff-based output representation. PIE~\cite{DBLP:journals/corr/abs-2302-07867} is a recent benchmarks that explore using LLMs for improving code performance. It evaluates various prompting methods and shows that these methods can significantly speed up code execution. Different from the above work that typically focuses on directly generating optimized code, \tool aims at iterative refining and boosting the initial results directly generated by LLMs in a search-based manner.

\subsection{Large Language Models for Software Engineering}
Recently, the advent of LLMs has significantly advanced various software engineering tasks~\cite{nijkamp2022codegen,DBLP:journals/corr/abs-2305-06161,DBLP:journals/tosem/GaoGHZNXL23}. 
A lot of research is dedicated to effectively harnessing the capability of LLMs by fine-tuning or prompt engineering for software engineering tasks~\cite{DBLP:conf/icml/ShrivastavaLT23,DBLP:conf/icse/GaoMG000L24,DBLP:conf/icse/YeM24}. For example, WizardCoder~\cite{DBLP:journals/corr/abs-2306-08568} fine-tunes LLMs with complex instructions for code generation. Xia et al.~\cite{DBLP:journals/corr/abs-2210-14179} leverage LLMs for program repair by using the cloze-stype prediction. TypeGEN~\cite{DBLP:conf/kbse/PengWWGL23} prompts LLMs with static analysis results and COT prompts for type inference. 
CHATRepair~\cite{DBLP:journals/corr/abs-2304-00385} iteratively evaluates programs on test cases and feeds the error messages to LLMs for further patch generation. Self-edit~\cite{DBLP:conf/acl/ZhangLLLJ23} utilizes compiler error messages to enhance the correctness of code generation. 

\subsection{SBSE and Large Language Models}
CodaMOSA~\cite{DBLP:conf/icse/LemieuxILS23} leverages LLMs to provide example test cases for under-covered functions when search-based testing hits a coverage stall. Tawosi et al.~\cite{DBLP:conf/ssbse/TawosiAL23} use available search-based methods to optimize the number and combination of few-shot examples for LLMs in the story point estimation task. Brownlee et al.~\cite{DBLP:conf/ssbse/BrownleeCEGHPSS23} introduce a method that employs LLMs as mutation operators for genetic improvement. Similarly, Kang and Yoo~\cite{DBLP:conf/gi-ws/KangY23} propose to leverage the capabilities of LLMs in code comprehension and generation for creating objective-tailored mutants. Dakhama et al.~\cite{DBLP:conf/ssbse/DakhamaELMP23} improve search-based fuzzing by using ChatGPT to parameterise C programs for gem5 testing. 

\section{Conclusion and future work}\label{sec:conclusion}
In this paper, we propose \tool, a search-based LLMs framework for code optimization. \tool synergistically integrates LLMs with evolutionary search, encompassing three components: an execution-based representative sample selection mechanism, an adaptive optimization pattern retrieval method, and a genetic operator-inspired chain-of-thought prompting method. 
Extensive experiments on four popular LLMs show that \tool can effectively guide LLMs towards identifying efficient optimization methods in the vast search space. 
In the future, we plan to apply our search-based LLMs framework to other tasks in software engineering such as program repair. 
Our source code and detailed experimental results are available at~\cite{SBLLM}. 



\bibliographystyle{IEEEtran}
\bibliography{sample-base.bib}

\end{document}